\documentclass[twocolumn]{aastex701}
\usepackage{amsmath}
\usepackage{multirow} 
\usepackage{soul}

\begin{document}

\title{Scientific Validation of the SPARC4 Pipeline: Multi-band Imaging, Polarimetry, and Photometric Time Series for Improved Characterization of Transiting Exoplanets}

\author[0000-0002-5084-168X]{Eder Martioli}
%\altaffiliation{Research Scientist}
\affiliation{Laborat\'{o}rio Nacional de Astrof\'{i}sica, Rua Estados Unidos 154, 37504-364, Itajub\'{a} - MG, Brazil}
\email{emartioli@lna.br}
\correspondingauthor{Eder Martioli}
\email{emartioli@lna.br}

\author[0000-0002-9459-043X]{Claudia V. Rodrigues}
%\altaffiliation{Research Scientist}
\affiliation{Instituto Nacional de Pesquisas Espaciais, Avenida dos Astronautas, 1.758, Jardim Granja, São José dos Campos, São Paulo, Brazil}
\email{claudia.rodrigues@inpe.br}

\author[0000-0002-2075-2424]{Julio C. N. Campagnolo }
\affiliation{Centro Federal de Educação Tecnológica Celso Suckow da Fonseca, Rio de Janeiro, Rio de Janeiro, Brazil}
\email{juliocampagnolo@gmail.com}

\author[0000-0002-0386-2306]{Francisco J. Jablonski}
\affiliation{Instituto Nacional de Pesquisas Espaciais, Avenida dos Astronautas, 1.758, Jardim Granja, São José dos Campos, São Paulo, Brazil}
\email{francisco.jablonski@inpe.br}

\author[0000-0002-0337-1363]{Ana Carolina Mattiuci}
\affiliation{Instituto Nacional de Pesquisas Espaciais, Avenida dos Astronautas, 1.758, Jardim Granja, São José dos Campos, São Paulo, Brazil}
\email{carol.mattiuci@gmail.com}

\author[0000-0002-8646-218X]{Fernando Falkenberg}
\affiliation{Instituto Nacional de Pesquisas Espaciais, Avenida dos Astronautas, 1.758, Jardim Granja, São José dos Campos, São Paulo, Brazil}
\email{fernando.marques@inpe.br}

\author[0009-0005-2307-4605]{Gustavo H. S. Santos}
%\author[0009-0005-2307-4605]{Gustavo Henrique da Silva Santos}
\affiliation{Instituto Nacional de Pesquisas Espaciais, Avenida dos Astronautas, 1.758, Jardim Granja, São José dos Campos, São Paulo, Brazil}
\email{gustavo.santos@inpe.br}

\author[0000-0001-7346-7638]{Marina M. C. Mello}
\affiliation{Instituto Nacional de Pesquisas Espaciais, Avenida dos Astronautas, 1.758, Jardim Granja, São José dos Campos, São Paulo, Brazil}
\email{marina.mello@inpe.br}

\author[0000-0001-6013-1772]{Isabel J. Lima}
\affiliation{Universidade Estadual Paulista, UNESP, Campus of Guaratinguetá, Av. Dr. Ariberto Pereira da Cunha, 333 - Pedregulho, 12516-410, Guaratinguetá - SP, Brazil}
\email{isabellima01@gmail.com}

\author[0000-0003-1549-4587]{Filipe V. M. Monteiro}
\affiliation{Universidade Estadual Paulista, UNESP, Campus of Guaratinguetá, Av. Dr. Ariberto Pereira da Cunha, 333 - Pedregulho, 12516-410, Guaratinguetá - SP, Brazil}
\email{filipe.monteiro@unesp.br}

\author[0000-0003-0680-1979]{Luciano Fraga}
\affiliation{Laborat\'{o}rio Nacional de Astrof\'{i}sica, Rua Estados Unidos 154, 37504-364, Itajub\'{a} - MG, Brazil}
\email{lfraga@lna.br}

\author[0000-0001-8179-1147]{Leandro de Almeida}
\affiliation{Instituto Nacional de Pesquisas Espaciais, Avenida dos Astronautas, 1.758, Jardim Granja, São José dos Campos, São Paulo, Brazil}
\affiliation{NSF's NOIRLab/SOAR Telescope, Casilla 603, La Serena 1700000, Chile}
\email{leandro.almeida@inpe.br}

\author[0000-0002-1387-2954]{Diego Lorenzo-Oliveira}
\affiliation{Laborat\'{o}rio Nacional de Astrof\'{i}sica, Rua Estados Unidos 154, 37504-364, Itajub\'{a} - MG, Brazil}
\email{dlorenzo@lna.br}

\author[0000-0002-0537-4146]{H\'elio D. Perottoni}
\affiliation{Observat\'{o}rio Nacional, MCTI, Rua Gal. Jos\'{e} Cristino 77, Rio de Janeiro, 20921-400, RJ, Brazil}
\email{hperottoni@on.br}

\author[0000-0001-9200-3441]{Laerte Andrade}
\affiliation{Laborat\'{o}rio Nacional de Astrof\'{i}sica, Rua Estados Unidos 154, 37504-364, Itajub\'{a} - MG, Brazil}
\email{landrade@lna.br}

\author[0000-0002-7095-4147]{Wagner Schlindwein}
\affiliation{Departamento de Física, Universidade Estadual de Maringá, Avenida Colombo, 5790, Maringá-PR, Brazil}
\affiliation{Instituto Nacional de Pesquisas Espaciais, Avenida dos Astronautas, 1.758, Jardim Granja, São José dos Campos, São Paulo, Brazil}
\email{wagner.schlindwein@inpe.br}

\author[0000-0001-7148-6886]{Denis Bernardes}
\affiliation{Laborat\'{o}rio Nacional de Astrof\'{i}sica, Rua Estados Unidos 154, 37504-364, Itajub\'{a} - MG, Brazil}
\email{dbernardes@lna.br}

%\collaboration{all}{(SPARC4 Pipeline)}

\begin{abstract}

High-cadence multi-band imaging and polarimetry have important scientific applications in astronomy. Observations of transits of exoplanets consist of a particular application that requires robust data reduction and analysis. We present the SPARC4 Pipeline, a suite of routines developed to process photometric and polarimetric data obtained with the instrument SPARC4 installed on the 1.6~m telescope at Pico dos Dias Observatory, Brazil. The scientific data products, up to the generation of high-cadence time series, are demonstrated using observations of several transiting exoplanetary systems in both photometric and polarimetric modes. These observations are used to produce stacked calibrated images, yielding sub-arcsecond astrometric accuracy even in sparse fields. The time series of these fields enabled a photometric characterization of the instrument. Observations of polarimetric standard stars yield an instrumental polarization below $0.06$\% and a linear polarization accuracy of $0.2$\%. Furthermore, transit observations of seven exoplanets with host-star magnitudes in the range $10.2 < V < 13.9$ demonstrate that SPARC4 achieves an average photometric precision of 0.02\% for a 15-minute cadence and a polarimetric precision of $\sim$0.02\% over hours-long time series. Finally, we jointly model the SPARC4 light curves together with TESS data (or K2 data in the case of HATS-9) using a Bayesian MCMC framework to refine the constraints on the physical parameters of the exoplanets, enabling a more accurate determination of the orbital periods and planetary radii, thereby providing improved constraints on the orbital and physical parameters of these hot Jupiters.

\end{abstract}

%% PASP uses Unified Astronomy Thesaurus (UAT) concepts:


\section{Introduction} \label{sec:Introduction}

The scientific capabilities and productivity of an astronomical instrument are greatly enhanced by employing a homogeneous instrumental calibration and a robust data reduction process, especially when performed by an automated pipeline. Each instrument has unique characteristics, requiring tailored calibration and reduction processes. The Simultaneous Polarimetry And Rapid Camera in 4 bands \citep[SPARC4;][]{Rodrigues2012,Rodrigues2024,Bernardes2025b} is a new instrument installed on the 1.6~m Perkin Elmer telescope at Pico dos Dias Observatory (OPD), Brazil. SPARC4 is an imager and polarimeter with a field of view of $5.7\times5.7$~arcmin$^{2}$, operating simultaneously in the four SDSS bands \citep[g, r, i, z;][]{Fukugita1996}, allowing for the acquisition of photometric and polarimetric time series at a high cadence.  SPARC4 had its first light in November 2022, underwent a commissioning and science verification in 2023, and initiated regular operations in March 2024. Table \ref{tab:sparc4instrumentparameters} summarizes the relevant SPARC4 instrument parameters and the readout time and noise for the avaliable reading modes of the detectors.

\begin{deluxetable*}{ccc}[bhpt]
\tablecaption{Summary of the SPARC4 instrument parameters.}
\label{tab:sparc4instrumentparameters}
\tablehead{\colhead{Parameter}  & \colhead{Value} & \colhead{Comment} }
\startdata
Number of imaging channels   & 4 & Bands similar to the SDSS $g, r, i, z$ \\
Polarimetric optics - analyzer & Savart prism & achromatic beam splitter \\
Polarimetric optics - retarder & $\lambda/2$ or $\lambda/4$ & Half- and quarter-waveplate \\
Focal ratio & $f/5$ & Optical speed at the focal plane \\ 
CCD detector size  (pixel $\times$ pixel) & $1024\times 1024$ & Squared field; full frame \\
Pixel size ($\mu$m $\times$ $\mu$m) & $13\times 13$ &  Squared pixels\\
Field-of-view  (arcmin $\times$ arcmin) & $5.7\times 5.7$ & Squared field; full frame \\
Plate scale   (arcsec per pixel) & 0.34 &  Binning $1\times 1$ \\
Maximum cadence  (frames per second) & 26 & Full frame; read rate of 30~MHz\\
Read time (s)\tablenotemark{a} / noise (e-/pixel)\tablenotemark{b} & 10.93/3.4 & Full frame; read rate of 0.1~MHz; Conv.\tablenotemark{c} \\
Read time (s)\tablenotemark{a} / noise (e-/pixel)\tablenotemark{b} & 1.11/4.8 & Full frame; read rate of 1~MHz; Conv.\tablenotemark{c} \\
Read time (s)\tablenotemark{a} / noise (e-/pixel)\tablenotemark{b} & 0.11/77.5 & Full frame; read rate of 10~MHz; EM\tablenotemark{c} \\
Read time (s)\tablenotemark{a} / noise (e-/pixel)\tablenotemark{b} & 0.06/141 & Full frame; read rate of 20~MHz; EM\tablenotemark{c} \\
Read time (s)\tablenotemark{a} / noise (e-/pixel)\tablenotemark{b} & 0.04/197 & Full frame; read rate of 30~MHz; EM\tablenotemark{c} \\
\enddata
\tablenotetext{a}{In frame-transfer (FT) mode, CCD readout introduces overhead only when the exposure time exceeds the read time. An extensive discussion about the instrument overheads is presented in \citet{Bernardes2025b}. }
\tablenotetext{b}{This value corresponds to the noise level of the camera in channel 1. The noise varies by up to a few percent between the four SPARC4 CCD cameras.}
\tablenotetext{c}{Conv. stands for conventional reading mode, while EM means electron-multiplying mode.}
\end{deluxetable*} 

The SPARC4 Pipeline \footnote{\url{https://github.com/edermartioli/sparc4-pipeline}} \citep{Martioli2025} is a suite of Python routines designed to reduce SPARC4 data for its main modes of operation. Namely, this covers regular imaging, dual-beam polarimetry of point-like sources using either a half-wave (L2; $\lambda/2$) or a quarter-wave (L4; $\lambda/4$) rotating retarder, as well as photometric and polarimetric time series.  We present results derived from the analysis of SPARC4 observations fully processed by the pipeline, demonstrating its performance and the quality of the resulting science products.

An important scientific application anticipated for SPARC4 is the characterization of transiting exoplanets through multi-band observations.  Transiting exoplanets play a crucial role in the contemporary era of planetary sciences, enabling the study of the physics of stars and their planets. Transits yield valuable information, including planet size and orbital distance \cite[e.g.,][]{Martioli2021}. They offer a unique opportunity to investigate the atmospheric composition from transmission and emission spectroscopy \citep[e.g.,][]{Seager2010,Snellen2018,Deming2019}, the composition and interior structure constrained by their bulk densities \citep{Dorn2015,Adibekyan2021}, and orbital obliquity measured by the Rossiter-McLaughlin effect \citep[e.g.,][]{Ohta2005,Martioli2020}.  The four-band simultaneous acquisition of SPARC4 aids in detecting and mitigating atmospheric effects affecting relative photometry, including differential refraction and red-noise systematics. In addition, the multi-band capability enables limb darkening and spot-crossing analyses \citep{Valio2025}, improving our understanding of stellar activity and providing a more precise transit depth determination. 

To this end, we observed the transits of exoplanets HATS-9~b, HATS-21~b, HATS-23~b, and HATS-24~b in photometric mode, and WASP-78~b, WASP-111~b, and WASP-123~b in polarimetric mode. We show how SPARC4 data can refine system parameters and enable studies of transit timing variations and other physical aspects of the system, such as wavelength-dependent stellar limb-darkening and spectral variations in planetary radius caused by atmospheric absorbers. 

This paper is organized as follows. Section~\ref{sec:observations} describes the observations obtained with SPARC4 and from the TESS and K2 missions. Section~\ref{sec:sparc4-pipeline} introduces the reduction methods employed by the SPARC4 Pipeline and an analysis of the pipeline products. Section~\ref{sec:transit-analysis} presents the analysis of exoplanet transits combining photometric SPARC4 and TESS/K2 data. Section~\ref{sec:polar-analysis} reports the polarimetric time series of exoplanet transits. Finally, Section \ref{sec:conclusions} provides the conclusions.

%--------------------------------------------------------------------
\section{Observations}
\label{sec:observations}

In this section, we report the observations obtained with SPARC4 at OPD, including both calibration and science data. We also describe the data collected from NASA’s Transiting Exoplanet Survey Satellite \citep[TESS,][]{tess_paper} and Kepler K2 \citep{kepler_k2_paper} missions, which are incorporated into our analysis of the exoplanet transits.

\subsection{SPARC4}
\label{sec:obs-sparc4}

The SPARC4 observations were conducted during an instrument commissioning night on 2023-11-07, an engineering night on 2024-07-18, and on other six nights allocated for the programs number OP2024A-004 and OP2024B-007 (PI: E. Martioli).   Table \ref{tab:sparc4observations} presents the log of SPARC4 observations that delivered the data presented in this paper.  
 
For each night, we obtained a set of daytime calibration frames, including bias and dome flats. Typically, we acquired 300 bias frames for each detector mode. We also took a sequence of flat-field exposures of the dome flat screen, following the recommended exposure times and lamp intensities published on the SPARC4 Information for Users website\footnote{\url{https://coast.lna.br/home/sparc4/information-for-users}}. During the commissioning night on 2023-11-07, flats were obtained only in photometric mode. For the other two nights with polarimetric data (2024-06-17 and 2024-09-08), we acquired flats in polarimetric mode for each position angle of the rotating waveplate used in the scientific observations. As we will demonstrate in Section~\ref{sec:polar-analysis}, this approach is recommended to achieve improved photometric precision.

At the beginning of each observation, we point the telescope to the exoplanet host star and adjust the field centering to include as many reference stars as possible. When selecting the field, we prioritize stars with brightness similar to the target and maximize the number of suitable reference stars within the field of view. Once the field is properly positioned, we set the telescope focus and begin observations. %acquired a series of exposures at different focus positions using the FOCUS tab in the S4GUI system \citep{Bernardes2025b}. To measure the optimal focus, we then used the tool \texttt{sparc4\_focus.py}, which is distributed within the SPARC4 Pipeline. This tool automatically detects point-like sources in the four-channel images and calculates an average full width at half maximum (FWHM) for each focus position. It subsequently fits a hyperbolic function to the focus data and determines the optimal focus position corresponding to the minimum FWHM value. Finally, we select a guide star and initiate auto-guiding.
We then take a test exposure using the same exposure time for all channels and measure the counts for the target as well as for the best reference stars in the field. Based on these measurements, we calculate an optimal exposure time for each channel, aiming for maximum counts to be approximately two-thirds of the saturation level. We also consider exposure times for the channels that share a reasonably small least common multiple, allowing us to set a number of exposures in each channel that fits within a cycle without introducing overhead. 

Then, we initiate a long series of cycles to cover the entire pre- and post-transit phases of the event. During the series, we monitor guiding performance, seeing conditions, and the counts in all channels. Occasionally, we interrupt the sequence to adjust exposure times, either to improve the signal-to-noise ratio (S/N) or to avoid saturation. The most frequently used exposure times are listed in Table~\ref{tab:sparc4observations}.

We observed three transits in the linear polarization mode with a rotating half-wave plate (L2). The observing procedure was the same as described above, except that within each cycle we acquired a sequence of exposures at different waveplate positions — ranging from positions 1 to 16, corresponding to angles from $0\degr$ to $337.5\degr$ in $22.5\degr$ steps. In addition to exoplanet transits, we also observed polarimetric standard stars, as listed in Table \ref{tab:sparc4observations}. 

\begin{deluxetable*}{cccccccc}
\tablecaption{Log of SPARC4 observations.}
\label{tab:sparc4observations}
\tablehead{\colhead{Object ID} & \colhead{V\tablenotemark{a}}  & \colhead{Obs. Type}  & \colhead{Date\tablenotemark{b}}      & \colhead{Duration}  & \colhead{Exp. Times\tablenotemark{c}} &  \colhead{\# Exposures\tablenotemark{c}} & \colhead{Inst. Mode} \\
\colhead{} & \colhead{(mag)}  & \colhead{}  & \colhead{} & \colhead{(h)} & \colhead{(s)} & \colhead{} & \colhead{}
 }
\startdata
%WD 2039-202& polar standard & 2023-11-07 & -      & 30,\,10,\,30,\,30                 & 8,\,24,\,8,\,8               & POLAR \\
WASP-78   & 11.96 & science & 2023-11-07 & 6.55      & 30,\,6,\,6,\,10                 & 690,\,3450,\,3450,\,2070               & POLAR \\
HATS-23   & 13.90 & science & 2024-05-05 & 4.25      & 30,\,15,\,15,\,30                 & 512,\,1021,\,1021,\,512                & PHOT \\
HATS-24   & 12.83 & science & 2024-06-05 & 4.86      & 40,\,8,\,10,\,20                 & 552,\,2748,\,1677,\,840                & PHOT \\
HATS-9    & 13.28 & science & 2024-06-15 & 7.40      & 20,\,8,\,8,\,20                 & 1281,\,3204,\,3204,\,1281                & PHOT \\
HATS-9    & 13.28 & science & 2024-06-17 & 7.81      & 20,\,8,\,8,\,20                 & 1353,\,3381,\,3381,\,1353                & PHOT \\
Hilt 652  & 10.61 & polar. standard & 2024-06-17 & -      & 20,\,4,\,4,\,10            & 34,\,162,\,162,\,66                & POLAR \\
WASP-123  & 11.06 & science & 2024-07-18 & 2.89      & 8,\,2,\,2,\,8                & 992,\,3968,\,3968,\,992              & POLAR \\
HATS-21   & 12.19 & science & 2024-09-06 & 3.19      & 12,\,4,\,4,\,6                 & 477,\,1433,\,1433,\,955                & PHOT \\
Hilt 715  & 9.56 & polar. standard & 2024-09-08 & -      & 10,\,2,\,1.5,\,0.8               & 16,\,64,\,80,\,112                & POLAR  \\
WASP-111  & 10.25 &  science & 2024-09-08 & 7.99      & 10,\,2,\,2,\,2.5              & 2179,\,10883,\,10883,\,8755                & POLAR \\
HD 13588  & 7.9 & polar. standard & 2024-09-08 & -      & 2,\,0.5,\,0.5,\,1               & 16,\,16,\,16,\,16                & POLAR\\
\enddata
\tablenotetext{a}{V magnitudes in the Johnson–Cousins photometric system were obtained from SIMBAD \citep{Wenger2000}.}
\tablenotetext{b}{Dates refer to the calendar date at the start of each observing night.}
\tablenotetext{c}{Exposure times are modal values. Exposure times and number of exposures are given as sets of four values, corresponding respectively to the SPARC4 bands $g$, $r$, $i$, and $z$.}
\end{deluxetable*} 

\subsection{TESS and K2}
\label{sec:obs-tess}

For each transiting planet, we retrieved the available TESS photometry time series from the Mikulski Archive for Space Telescopes (MAST)\footnote{\url{mast.stsci.edu}}. One of our targets, HATS-9, was not observed by TESS, but it was observed by the Kepler K2 mission, whose data we use for our analysis instead.  Table~\ref{tab:tessobservations} presents the log of TESS and Kepler observations that we used in our analysis.  

\begin{deluxetable*}{ccccccc}
\tablecaption{Log of TESS and K2 observations of the exoplanet systems.}
\label{tab:tessobservations}
\tablehead{
\colhead{Object ID} & \colhead{Object} & \colhead{Mission \&} & \colhead{TSTART} & \colhead{TSTOP} & \colhead{Duration} & \colhead{Sampling} \\
\colhead{} & \colhead{Mission ID}  & \colhead{Sector}  & \colhead{(UTC)} &  \colhead{(UTC)} & \colhead{(d)} & \colhead{(minutes)}
}
\startdata
 WASP-78 & TOI-449 & TESS S5 & 2018-11-15T11:47:43.531 & 2018-12-11T14:51:12.613 & 26.1 & 2.0 \\
 WASP-78 & TOI-449 & TESS S31 & 2020-10-22T00:24:37.623 & 2020-11-16T10:45:15.297 & 25.4 & 2.0 \\
 WASP-78 & TOI-449 & TESS S32 & 2020-11-20T17:25:15.046 & 2020-12-16T17:26:27.516 & 26.0 & 2.0 \\
\hline
 HATS-23 & TOI-1065 & TESS S13 & 2019-06-19T10:34:42.652 & 2019-07-17T20:40:43.863 & 28.4 & 2.0 \\
 HATS-23 & TOI-1065 & TESS S67 & 2023-07-01T03:30:02.359 & 2023-07-28T21:35:23.788 & 27.8 & 2.0 \\
\hline
 HATS-24 & TOI-1084 & TESS S13 & 2019-06-19T10:10:05.361 & 2019-07-17T20:39:26.260 & 28.4 & 2.0 \\
 HATS-24 & TOI-1084 & TESS S39 & 2021-05-27T06:35:05.896 & 2021-06-24T05:21:18.012 & 27.9 & 0.3 \\
 HATS-24 & TOI-1084 & TESS S66 & 2023-06-02T04:14:52.517 & 2023-06-30T22:21:07.191 & 28.8 & 2.0 \\
\hline
HATS-9 & K2-142 & Kepler K2 & 2015-10-04T17:52:39.907 & 2015-12-26T08:35:28.392 & 82.6 & 30.0 \\ 
\hline
 WASP-123 & TOI-1069 & TESS S13 & 2019-06-20T12:19:19.310 & 2019-07-17T20:41:37.891 & 27.3 & 2.0 \\
 WASP-123 & TOI-1069 & TESS S27 & 2020-07-05T18:47:28.894 & 2020-07-30T03:30:55.799 & 24.4 & 2.0 \\
 WASP-123 & TOI-1069 & TESS S67 & 2023-07-01T03:30:46.078 & 2023-07-28T21:36:22.526 & 27.8 & 2.0 \\
\hline
 HATS-21 & TOI-1071 & TESS S13 & 2019-06-19T10:10:16.588 & 2019-07-17T20:39:59.936 & 28.4 & 2.0 \\
 HATS-21 & TOI-1071 & TESS S66 & 2023-06-02T04:14:48.242 & 2023-06-30T22:21:27.877 & 28.8 & 2.0 \\
\hline
 WASP-111 & TOI-143 & TESS S1 & 2018-07-25T19:09:27.322 & 2018-08-22T15:59:53.101 & 27.9 & 2.0 \\
 WASP-111 & TOI-143 & TESS S28 & 2020-07-31T08:31:12.943 & 2020-08-25T14:27:19.090 & 25.2 & 2.0 \\
 WASP-111 & TOI-143 & TESS S68 & 2023-07-29T02:46:24.365 & 2023-08-25T15:22:39.147 & 27.5 & 2.0 \\
\enddata
\end{deluxetable*} 

We inspected these light curve data and analyzed the transits using the methods described in \cite{Martioli2021}, resulting in well-constrained models for the system parameters. In particular, the TESS and Kepler data provide precise measurements of the orbital period and transit times, allowing us to accurately predict the ephemerides of the transits observed with SPARC4. 

%--------------------------------------------------------------------
\section{The SPARC4 Pipeline}
\label{sec:sparc4-pipeline}

The SPARC4 Pipeline is a suite of Python routines designed to reduce SPARC4 data. The code within the pipeline framework is organized as follows. A main script reads the reduction parameters from a \texttt{yaml} file and executes all the reduction tasks, relying on a collection of functions grouped into libraries. The data reduction algorithms are primarily implemented through third-party packages. The main packages include \texttt{ASTROPOP} \citep{Campagnolo2019}, \texttt{PHOTUTILS} \citep{larry_bradley_2024_13989456}, \texttt{ASTROPY} \citep{2013A&A...558A..33A,2018AJ....156..123A,2022ApJ...935..167A}, \texttt{SCIPY}, \texttt{NUMPY} \citep{VanDerWalt2011}, \texttt{MATPLOTLIB} \citep{Hunter2007}, and \texttt{ASTROQUERY} \citep{Ginsburg2019}.
%--------------------------------------------------------------------
\subsection{Overview of SPARC4 data reduction}
\label{sec:data-reduction}

The pipeline processes the data for each SPARC4 channel independently. The first step consists of creating a database in a CSV file (\texttt{*db.csv}), which is a table containing file paths in the first column and several other columns with information extracted from the header of raw images. This information allows the identification of reduction groups. For calibration data (bias and flats), the groups are selected based on frames that match specific instrumental and detector modes. Similarly, the science frames are grouped according to matching detector mode and instrument mode (PHOT or POLAR), as well as by object name.

For each reduction group, the pipeline uses the \texttt{ASTROPOP}\footnote{\url{https://github.com/juliotux/astropop/tree/main/astropop}} package \citep{Campagnolo2019,2018ascl.soft05024C} to combine zero (bias) frames by taking the median of raw frames and applies gain correction to create a master zero frame in units of electrons. It combines flat-field frames in a similar manner and normalizes the result by the mean flux to produce a normalized master flat-field.  In POLAR mode, the pipeline also computes a separate master flat for each waveplate position. The results are stored in the Master Calibration products, which are described in Appendix~\ref{sec:mastercalibproducts}.

Then, whether in polarimetry or photometry mode, a given reduction group of science data is processed as follows. For each science frame, we apply gain correction, zero subtraction and division by flat-field. For image registration, the frames are aligned using a cross-correlation algorithm that applies linear shifts in pixel space. A subset of calibrated registered science frames is selected for the calculation of a sigma-clipped mean stack, using nearest-neighbor interpolation.  The stack is used to find point-like sources, match with an external reference catalog, solve astrometry and save the astrometric solution to WCS header keywords. The \texttt{PHOT\_THRESHOLD} parameter specifies the S/N detection threshold for identifying sources, while the \texttt{PHOT\_APERTURES} parameter defines a set of aperture radii. Aperture photometry is then carried out for all detected sources across the defined apertures, and the resulting photometric data are stored in catalog table extensions corresponding to each aperture. The calibrated stacked image is saved along with all catalogs in a multi-extension FITS file (\texttt{*stack.fits}, see Appendix~\ref{sec:sciimgproducts}).  In polarimetric mode, the source images are duplicated. Then, the detected sources are matched in pairs using the algorithm of \cite{Campagnolo2019} and two independent catalogs are created, one for each polarimetric beam. The sources with unmatched pairs are not included in the catalogs.  

Next, the pipeline reduces each science frame individually. It applies zero, flat, and gain corrections, registers the image through cross-correlation with the stack, applies the calculated offset to the astrometric solution, and saves it to WCS in the header. The source catalog from the stacked image is used to perform multi-aperture photometry on the same sources in each individual frame by applying the measured registration offsets to their pixel coordinates. This ensures that all catalogs have identical formats and contain the same sources as the stacked image. This consistency is particularly important for robust and reliable reduction of long time series observations, even under variable weather conditions. Each processed frame and photometric catalogs are saved into a FITS file (\texttt{*proc.fits}, see Appendix~\ref{sec:sciimgproducts}).

From this point onward, the data reduction for observations obtained in PHOT or POLAR modes follows different paths. In PHOT mode, a time series of all photometric quantities (see Appendix~\ref{sec:sciimgproducts}) for all sources and apertures is generated and saved into a light curve FITS product (\texttt{*lc.fits}, see Appendix~\ref{sec:lcproduct}). In POLAR mode, whether in L2 or L4, the frames are grouped into sequences based on the waveplate position angle. Each sequence is treated as a group that can generate a polarimetric measurement. The SPARC4 instrument supports 16 waveplate position angles, and while it is common to obtain sequences in all positions, it is not mandatory. Moreover, since each channel may have different exposure times, the number of frames per waveplate position may also vary. The pipeline handles this, where it identifies and groups the data obtained in the same sequence to generate a single polarimetric measurement per sequence. All recognized sequences are processed, and the polarimetry is computed using the algorithms described in \cite{Campagnolo2019} and the results are saved into a polarimetry FITS product (\texttt{*polar.fits}, see Appendix~\ref{sec:polarproduct}). Finally, the pipeline combines all polarimetric products to create a time series of all polarimetric quantities, which is saved into a polar time series FITS product (\texttt{*ts.fits}, see Appendix~\ref{sec:tsproduct}).

The SPARC4 pipeline products are described in Appendix~\ref{app:s4products}. The following sections provide a more detailed description of specific methodologies adopted by the pipeline that are crucial for ensuring data reduction quality. 

\subsection{Astrometry}
\label{sec:astrometry}

The astrometric calibration in the SPARC4 Pipeline is performed on the stacked science image, which is assumed to be the highest-quality image for a given field observed during the night. The process begins by reading the WCS information from a set of four full-frame calibrated SPARC4 FITS images -- one for each channel -- all of which have well-calibrated WCS solutions. A sample of calibrated images is included in the SPARC4 pipeline package distribution in the \texttt{calibdb} directory.

The WCS solution obtained from the reference images is offset using the right ascension (RA) and declination (Dec) equatorial coordinates found in the image header, which are provided by the telescope control system (TCS). These coordinates reflect the initial telescope pointing and are not updated when small offsets are applied to refine the field centering. As a result, they may differ from the actual field center by up to a few arcminutes.

An online catalog is queried over a region larger than the SPARC4 field of view (FoV), by default 1.5 times the nominal FoV, though this factor can be adjusted by the user. The catalog query is performed using \texttt{Astroquery} tools. Users can query catalogs available on VizieR \citep{Ochsenbein2000}, but the default and most reliable option is to use the Gaia DR3 catalog, accessed via the tool \texttt{astroquery.gaia.Gaia}.   

\begin{figure*}
   \centering
       \includegraphics[width=1.0\hsize]{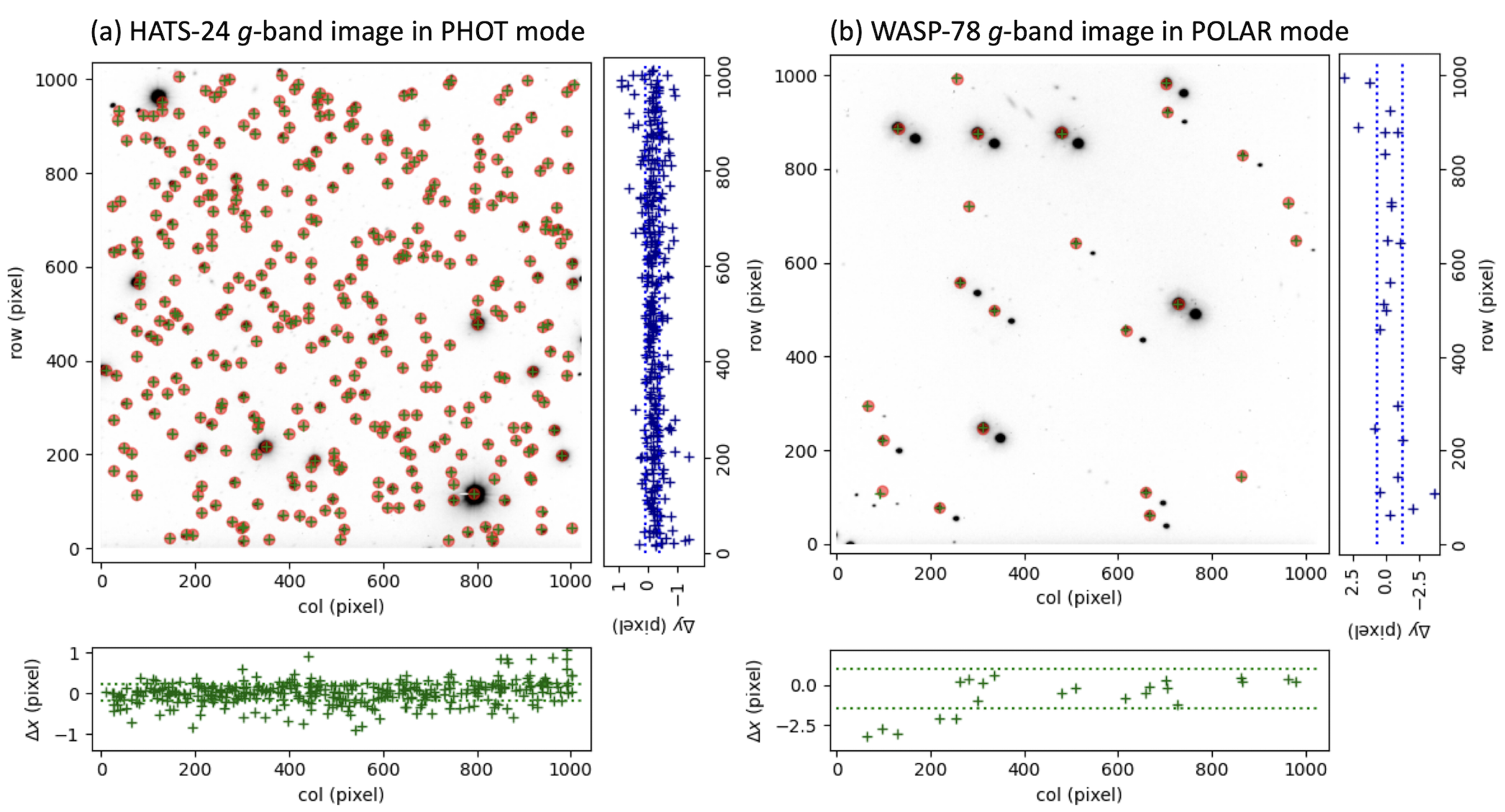}
      \caption{ Stacked images of the fields of HATS-24 (a) and WASP-78 (b) observed with SPARC4 in PHOT and POLAR modes, respectively, both in channel 1 (\textit{g} band). Red circles indicate the Gaia DR3 catalog coordinates, transformed into pixel coordinates using the WCS solution calculated by the pipeline. The crosses represent the source positions detected by the pipeline. The panels in the bottom and right-hand side display the differences between these two sets of pixel coordinates in the x- and y-directions, respectively. The dotted lines indicate the $\pm1\sigma$ dispersion, calculated as the standard deviation of the residuals, yielding $\sigma=0.24$~pixel ($0.08$~arcsec) for the HATS-24 field and $\sigma=1.2$~pixel ($0.4$~arcsec) for the WASP-78 field, based on 357 and 23 matched sources, respectively.
      }
      \label{fig:s4pipe-astrometry}
  \end{figure*}

The source lists obtained from the pipeline source detection and from the astrometric reference catalog are both sorted by magnitude, from brightest to faintest. To match sources between the two catalogs, we use the \texttt{find\_transform} algorithm from \texttt{aafitrans}\footnote{\url{https://github.com/prajwel/aafitrans}}, a tool built on top of the \texttt{Astroalign} package \citep{Beroiz2020}. This algorithm returns the matched source pairs, which are then passed to the \texttt{Astropy} function \texttt{wcs.fit\_wcs\_from\_points} to compute a new WCS solution for the image. To optimize the astrometric solver, the pipeline first derives an initial solution without high-order distortions, then constrains the distortion order according to the number of matched sources to prevent overfitting. The updated WCS is saved to the image header. Figure~\ref{fig:s4pipe-astrometry} shows the stacked images of the HATS-24 and WASP-78 fields, observed in channel 1 (\textit{g} band), as examples in both PHOT and POLAR modes, illustrating the pixel coordinate differences across the field. Note in Figure~\ref{fig:s4pipe-astrometry}~b that the pipeline adopts a convention in which the WCS solution corresponds to the coordinate system of the northern polarimetric beam.

\begin{figure}
   \centering
      \includegraphics[width=8.5cm]{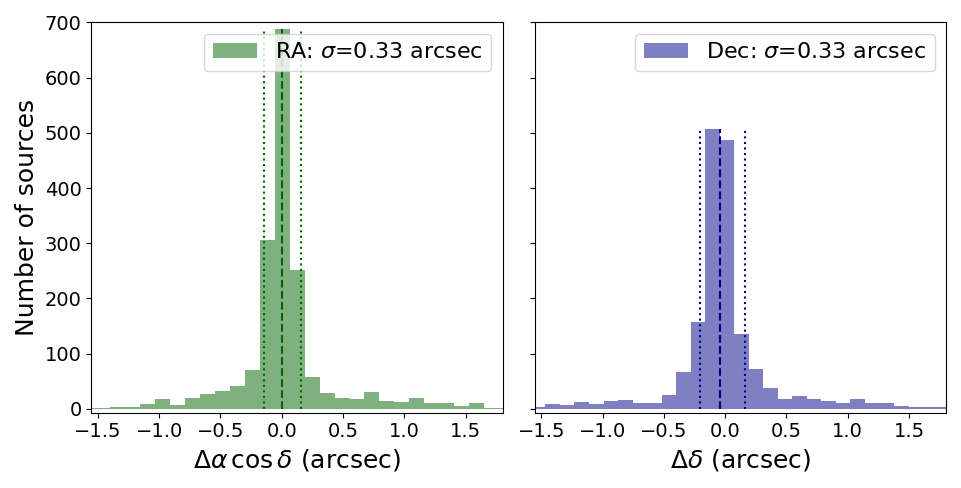}
      \caption{Distributions of $\Delta\alpha\cos{\delta}$ (left) and $\Delta\delta$ (right) for the \textit{g} band. The quantities $\Delta\alpha$ and $\Delta\delta$ represent the differences between the RA and Dec coordinates derived by the pipeline and those from the Gaia DR3 catalog, computed from all sources detected in the stacked images obtained during the eight nights of observations presented in this work. The dotted lines and values in the legends indicate the standard deviations of the distributions.
      }
      \label{fig:astrometry_quality}
  \end{figure}

The global astrometric accuracy achieved by the pipeline was evaluated from the distributions of the RA ($\alpha$) and Dec ($\delta$) coordinate differences, defined as $\Delta\alpha = \alpha_{\rm pipeline} - \alpha_{\rm Gaia\,DR3}$ and $\Delta\delta = \delta_{\rm pipeline} - \delta_{\rm Gaia\,DR3}$. These differences were computed using the WCS solutions derived by the pipeline ($\alpha_{\rm pipeline}$, $\delta_{\rm pipeline}$) and those from the Gaia DR3 catalog ($\alpha_{\rm Gaia\,DR3}$, $\delta_{\rm Gaia\,DR3}$). Figure~\ref{fig:astrometry_quality} shows the $\Delta\alpha\cos{\delta}$ and $\Delta\delta$ distributions for the \textit{g} band, measured from all sources detected in the stacked images obtained for the eight nights of observations presented in this work. The astrometric accuracy was estimated from the standard deviation of these distributions, yielding RA/Dec values of 0.22/0.35 arcsec for the \textit{g} band (1904 sources). Applying the same procedure to the other SPARC4 channels, we obtain: \textit{r} band -- 0.22/0.35 arcsec (2374 sources); \textit{i} band -- 0.15/0.15 arcsec (2604 sources); and \textit{z} band -- 0.22/0.31 arcsec (2728 sources).

Therefore, while one can achieve accuracy at the hundredths-of-an-arcsecond level for crowded fields, such as the HATS-24 field presented in Figure \ref{fig:s4pipe-astrometry}, the limiting accuracy provided by the pipeline products in sparse fields is around 0.3 arcseconds, which is considered satisfactory as it is sufficient for reliable source matching with external catalogs. While there is potential for improving the astrometric precision, especially for specific scientific applications that demand higher accuracy, such improvements are beyond the scope of the pipeline.

\subsection{Photometry}
\label{sec:photometry}

The pipeline calculates source fluxes and their associated errors using circular aperture photometry, performed by the \texttt{ApertureStats} function from the \texttt{Photutils} package \citep{larry_bradley_2024_13989456}. The flux is computed as the background-subtracted sum of electron counts divided by the exposure time, yielding units of electrons per second. The background flux is estimated as the median of a sigma-clipped set of pixels within an offset annulus, scaled by the area of the source aperture.  The background annulus dimensions are fixed, with inner and outer radii of 25 and 50 pixels (or 8.5 and 17 arcseconds), respectively. For large apertures ($r>23$~pixels), the pipeline automatically resizes the annulus to enforce an inner radius at least 2 pixels larger than the source aperture and a minimum annulus width of 10 pixels.   For each source, the pipeline calculates fluxes using a set of aperture radii defined in the parameters file. All flux values, $f$, are converted to instrumental magnitudes using the expression $m=-2.5\log(f)$. The magnitudes are saved in a FITS table, along with related quantities such as the RA, Dec, and pixel coordinates, aperture size, full width at half maximum (FWHM), background magnitudes, and a photometric quality flag. The FWHM is estimated in two ways: (i) by fitting a 2D Gaussian to the point-spread function (PSF) of each star in the stacked image, which is more accurate but slower, and (ii) using the analytic approximation 

$$FWHM = 2.355 \sqrt{\frac{f'}{2\pi f_{\rm peak}}}, $$

\noindent where $f'=\sum{f_i}$ is the integrated flux within a window of $25\times25$~pixels centered on the star, and $f_{\rm peak}$ is the maximum flux within the same window. The FITS tables containing the values of photometric quantities for all sources are stored in the catalog extensions of the reduced FITS images. 

Although the pipeline yields only instrumental magnitudes, we applied an absolute flux calibration procedure to the SPARC4 data processed by the pipeline. The results are presented in Appendix~\ref{app:photometricCalibration}.

%\subsection{Differential photometry}
%\label{sec:diffphot}

In many scientific applications, it is desirable to obtain time series of differential photometry to measure relative brightness variations. Differential photometry is calculated as the ratio between the flux of the target and the flux of reference stars, which are assumed to remain constant over time. The pipeline does not provide a differential photometry product, as the procedure is dependent on the specific science case. The pipeline provides the \texttt{plot\_light\_curve} tool to compute differential photometry from the time series products (\texttt{*lc.fits}). It returns the differential photometry relative to each comparison star, the combined flux of all selected comparison stars, and other relevant photometric quantities. All light curves of the transits of exoplanets presented in Section~\ref{sec:transit-analysis} have been obtained using this tool.

\subsection{Polarimetry}
\label{sec:polarimetry}

The SPARC4 instrument is designed to measure linear (Stokes Q and U) and circular (Stokes V) polarization of point-like sources using the dual-beam technique \citep[e.g.,][]{Magalhaes1984}. Therefore, SPARC4 employs a rotating retarder wave plate -- either a half-wave or quarter-wave plate -- and a Savart prism, which splits the incoming light into two orthogonally polarized beams \citep[see][]{Rodrigues2012,Rodrigues2024}. The pipeline implements the methodology and equations developed by \cite{Magalhaes1984} for half-wave plate and \cite{Rodrigues1998} for quarter-wave plate, following the format summarized by \cite{Campagnolo2019} and \cite{Mattiuci2024}. The polarimetric formalism is also described in the \texttt{ASTROPOP} documentation\footnote{\url{https://astropop.readthedocs.io/en/latest/reduction/polarimetry.html}}. 
The Stokes parameters are in flux units. In most cases, obtaining the polarization fractions is both scientifically sufficient and observationally more practical. For this reason, it is common to define the dimensionless normalized Stokes parameters $q = Q/I$, $u = U/I$, and $v = V/I$, where $I$ is the total flux.
The normalized Stokes parameters are calculated using a least-squares fit applied to the modulation of the normalized flux difference between the two orthogonal beams, $f_\parallel(\phi)$ and $f_\perp(\phi)$, defined as $z(\phi)$:
\begin{equation}
z(\phi) = \frac{f_\parallel(\phi) - k f_\perp(\phi)}{f_\parallel(\phi) + k f_\perp(\phi)},
\label{eq:zpol}
\end{equation}

\noindent where $\phi$ is the position angle of the retarder, and $k$ is a normalization factor related to the sensitivity of the two beams (see \citealt{Magalhaes1984} for half-wave plate and \citealt{Lima2021} for quarter-wave plate expressions).

For a half-wave plate (used for measuring linear polarization, through the normalized Stokes parameters $q$ and $u$), the following model is fitted:

\begin{equation}
z(\phi) = q \cos{4\phi} + u \sin{4\phi}.
\label{eq:zpolStokesQU}
\end{equation}

\noindent While for a quarter-wave plate (used for measuring both linear and circular polarization), the model also includes the normalized Stokes $v$:

\begin{equation}
z(\phi) = q \cos^{2}{2\phi} + u \sin{2\phi}\cos{2\phi} - v \sin{2\phi}.
\label{eq:zpolStokesQUV}
\end{equation}

%The pipeline computes the normalized Stokes parameters $q = Q/I$, $u = U/I$, and $v = V/I$ (where $I$ is the total light intensity), the degree of linear 
The pipeline computes the normalized Stokes parameters $q$, $u$, and $v$, the degree of linear polarization $p = \sqrt{q^2 + u^2}$, and the corresponding polarization angle $\theta = \tfrac{1}{2}\arctan\left(\tfrac{u}{q}\right)$ for all sources in the catalogs, and stores the results in a dedicated polarimetric data product (see Appendix~\ref{app:s4products}). Equations \ref{eq:zpolStokesQU} and \ref{eq:zpolStokesQUV} can be fitted using data from any number of frames taken at various wave plate position angles. To ensure reliable polarimetric measurements, a minimum of four (eight) non-redundant half-(quarter-)wave plate angles is required. However, it is considered good practice to obtain measurements at all 16 predefined wave-plate angles. Doing so enhances the robustness of the polarimetric results. The pipeline supports fitting the data using all 16 positions by default. However, it can also be configured to use a subset of positions chosen at regular intervals to increase temporal resolution (cadence), even when the full 16-position sequence has been observed.  

In Appendix~\ref{app:polarstandards}, Figures~\ref{fig:Hilt652_POLAR_L2}, \ref{fig:Hilt715_POLAR_L2}, and \ref{fig:HD13588_POLAR_L2} present the polarimetry results for the polarized standard stars Hilt~652 and Hilt~715, and the unpolarized standard star HD~13588, obtained with the pipeline in the four SPARC4 bands. A summary of the polarimetric results for these standard stars is given in Table~\ref{tab:polarcalibresults}. We also report the theoretical polarimetric error, $\sigma_{t} \propto 1/({\rm S/N})$, as calculated by the pipeline. This quantity allows for a direct comparison with the measurement uncertainties, which are of the same order of magnitude in the cases shown here and in general. These results confirm that the pipeline is working properly for polarimetric reduction and also demonstrate the high quality of the data. In this paper, the polarimetric standard stars are used primarily to validate the pipeline, though we also provide comparisons with values from the literature in the remainder of this section. A more detailed study of the SPARC4 polarimetric data, including an assessment of the long-term polarimetric stability of the instrument based on standard-star observations from the first two years of operation is currently underway and will be presented in a forthcoming dedicated publication (Mattiuci et al., in prep.). Preliminary results of this analysis reported in \cite{Mattiuci2024, Mattiuci2025} are consistent with the analysis presented here.

\begin{deluxetable*}{ccccccccc}
\tablecaption{Polarimetric measurements of two polarized standard stars and one unpolarized star obtained with the SPARC4 pipeline and literature values. The literature values are reported in the Johnson–Morgan photometric system, while SPARC4 uses the Sloan SDSS system. The reference polarization angle is taken as the mean of all reference values across all bands, yielding $\bar{\theta}_{\rm ref} = 49.62\pm0.04$~deg for Hilt~715 and $\bar{\theta}_{\rm ref} = 179.42\pm0.03$~deg for Hilt~652.}
\label{tab:polarcalibresults}
\tablehead{
\colhead{Object ID} & \colhead{Band} & \colhead{$q$ (\%)} & \colhead{$u$ (\%)} & \colhead{$p$ (\%)} & \colhead{$\sigma_{t}$(\%)} & \colhead{$\theta$ (deg)} & \colhead{$\theta-\bar{\theta}_{\rm ref}$ (deg)} & \colhead{Source}
}
\startdata                           
Hilt~715                    & $g$ & -4.34$\pm$0.04 & -3.89$\pm$0.04 & 5.83$\pm$0.04 & 0.030 & 110.91$\pm$0.17 & $61.29\pm0.18$ & SPARC4 (this work)  \\
                            & $r$ & -4.132$\pm$0.019 & -4.335$\pm$0.019 & 5.988$\pm$0.019 & 0.012 & 113.19$\pm$0.09 & $63.56\pm0.10$ & SPARC4 (this work)  \\
                            & $i$ & -3.479$\pm$0.025 & -4.004$\pm$0.025 & 5.304$\pm$0.025 & 0.015 & 114.51$\pm$0.13 & $64.88\pm0.14$ & SPARC4 (this work)  \\
                            & $z$ & -3.20$\pm$0.03 & -3.03$\pm$0.03 & 4.41$\pm$0.03 & 0.020 & 111.71$\pm$0.19 & $62.08\pm0.20$ & SPARC4 (this work) \\
                            & $B$ & $-0.992\pm0.024$ &  $5.715\pm0.024$ & $5.801\pm0.024$ & & $49.70\pm0.07$ & $0.08\pm0.08$ & FORS2 \citep{Cikota2017}  \\
                            & $B$ & $-0.98\pm0.03$ & $5.63\pm0.01$ & $5.71\pm0.01$ & & $49.93\pm0.15$ & $0.31\pm0.15$ & IPOL \citep{Fossati2007}  \\
                            & $B$ & $-0.97\pm0.01$ & $5.66\pm0.01$ & $5.74\pm0.01$ & & $49.88\pm0.07$ & $0.26\pm0.08$ & PMOS \citep{Fossati2007}  \\
                            & $V$ & $-0.955\pm0.012$ &  $6.021\pm0.012$ & $6.100\pm0.012$ & & $49.80\pm0.05$ & $0.18\pm0.06$ & FORS2 \citep{Cikota2017}   \\
                            & $V$ & $-0.99\pm0.02$ & $6.07\pm0.02$ & $6.15\pm0.02$ & & $49.62\pm0.08$ & $-0.00\pm0.09$ & IPOL \citep{Fossati2007}  \\
                            & $V$ & $-0.93\pm0.01$ & $6.06\pm0.01$ & $6.13\pm0.01$ & & $49.35\pm0.03$ & $-0.27\pm0.05$ & PMOS \citep{Fossati2007}  \\
                            & $R$ & $-0.864\pm0.005$ &  $5.753\pm0.006$ & $5.818\pm0.006$ & & $49.70\pm0.04$ & $0.08\pm0.05$ & FORS2 \citep{Cikota2017}  \\
                            & $R$ & $-0.98\pm0.05$ & $5.69\pm0.07$ & $5.77\pm0.07$ & & $49.90\pm0.24$ & $0.28\pm0.24$ & IPOL \citep{Fossati2007}   \\
                            & $I$ & $-0.796\pm0.006$ &  $4.926\pm0.006$ & $4.990\pm0.006$ & & $49.44\pm0.05$ & $-0.18\pm0.06$ & FORS2 \citep{Cikota2017}  \\
                            & $I$ & $-0.70\pm0.03$ & $5.07\pm0.01$ & $5.12\pm0.01$ & & $48.92\pm0.19$ & $-0.70\pm0.19$ & PMOS \citep{Fossati2007}  \\
\hline         
Hilt~652                    & $g$ & -3.16$\pm$0.04 & 4.96$\pm$0.04 & 5.88$\pm$0.04 & 0.017 & 61.23$\pm$0.18 & $-118.18\pm0.18$ & SPARC4 (this work)\\
                            & $r$ & -3.716$\pm$0.011 & 4.920$\pm$0.011 & 6.166$\pm$0.011 & 0.006 & 63.53$\pm$0.05 & $-115.88\pm0.06$ & SPARC4 (this work) \\
                            & $i$ & -3.688$\pm$0.015 & 4.343$\pm$0.015 & 5.698$\pm$0.015 & 0.008 & 65.17$\pm$0.07 & $-114.25\pm0.08$ & SPARC4 (this work) \\
                            & $z$ & -2.67$\pm$0.04 & 3.96$\pm$0.04 & 4.78$\pm$0.04 & 0.010 & 61.96$\pm$0.24 & $-117.45\pm0.25$ & SPARC4 (this work)  \\
                            & $B$ & $5.948\pm0.017$ &  $-0.054\pm0.017$ & $5.948\pm0.017$	& & $179.52\pm0.05$ & $0.11\pm0.06$ & FORS2 \citep{Cikota2017}  \\
                            & $B$ & $5.70\pm0.01$ & $-0.11\pm0.03$ & $5.70\pm0.01$	& & $179.47\pm0.13$ & $0.06\pm0.13$ & IPOL \citep{Fossati2007}   \\
                            & $B$ & $5.80\pm0.01$ & $-0.03\pm0.01$ & $5.80\pm0.01$	& & $179.85\pm0.05$ & $0.44+\pm0.06$ & PMOS \citep{Fossati2007}   \\                         
                            & $V$ &  $6.367\pm0.009$ & $-0.198\pm0.009$ & $6.371\pm0.009$ & & $179.44\pm0.03$ & $0.03\pm0.04$ & FORS2 \citep{Cikota2017}   \\
                            & $V$ & $6.24\pm0.03$ & $-0.18\pm0.04$ & $6.25\pm0.03$ & & $179.18\pm0.20$ & $-0.23\pm0.20$ & IPOL \citep{Fossati2007}  \\
                            & $V$ & $6.31\pm0.01$ & $-0.16\pm0.02$ & $6.32\pm0.01$ & & $179.27\pm0.07$ & $-0.14\pm0.08$ & PMOS \citep{Fossati2007}  \\
                            & $R$ & $6.214\pm0.004$ & $-0.218\pm0.004$ & $6.218\pm0.004$ & & $179.39\pm0.03$ & $-0.03\pm0.04$ & FORS2 \citep{Cikota2017}  \\
                            & $R$ & $6.07\pm0.02$  & $-0.13\pm0.02$ & $6.07\pm0.02$ & & $179.39\pm0.10$ & $-0.03\pm0.10$ & IPOL \citep{Fossati2007}   \\
                            & $I$ & $5.613\pm0.004$ & $-0.041\pm0.004$ & $5.613\pm0.004$ & & $179.46\pm0.03$ & $0.05\pm0.04$ & FORS2 \citep{Cikota2017}  \\
                            & $I$ & $5.61\pm0.04$  & $-0.16\pm0.02$ & $5.61\pm0.04$ &  &$179.18\pm0.11$ & $-0.23\pm0.11$ & PMOS \citep{Fossati2007}  \\
\hline                            
                            & $g$ & 0.047$\pm$0.023 & 0.008$\pm$0.024 & 0.048$\pm$0.024 & 0.015 & 4$\pm$14 & & SPARC4 (this work) \\
                            & $r$ & 0.048$\pm$0.030 & 0.024$\pm$0.030 & 0.054$\pm$0.030 & 0.017 & 13$\pm$16 & & SPARC4 (this work) \\
        HD~13588            & $i$ & 0.018$\pm$0.041 & -0.033$\pm$0.041 & 0.037$\pm$0.041 & 0.027 & 149$\pm$31 & & SPARC4 (this work) \\
        (unpol.)            & $z$ & 0.061$\pm$0.033 & 0.003$\pm$0.033 & 0.061$\pm$0.033 &0.025 & 1$\pm$15 & & SPARC4 (this work)  \\            
                            & $B$ &  &  & $0.100\pm0.008$	& & & & FORS2 \citep{Cikota2017}  \\
                            & $V$ &  &  & $0.105\pm0.007$ & & & & FORS2 \citep{Cikota2017}   \\
                            & $R$ &  &  & $0.114\pm0.004$ & & & & FORS2 \citep{Cikota2017}  \\
                            & $I$ &  &  & $0.126\pm0.005$ & & & & FORS2 \citep{Cikota2017}  \\                            
\enddata
\end{deluxetable*} 

\begin{figure*}
\centering
\includegraphics[width=1.\hsize]{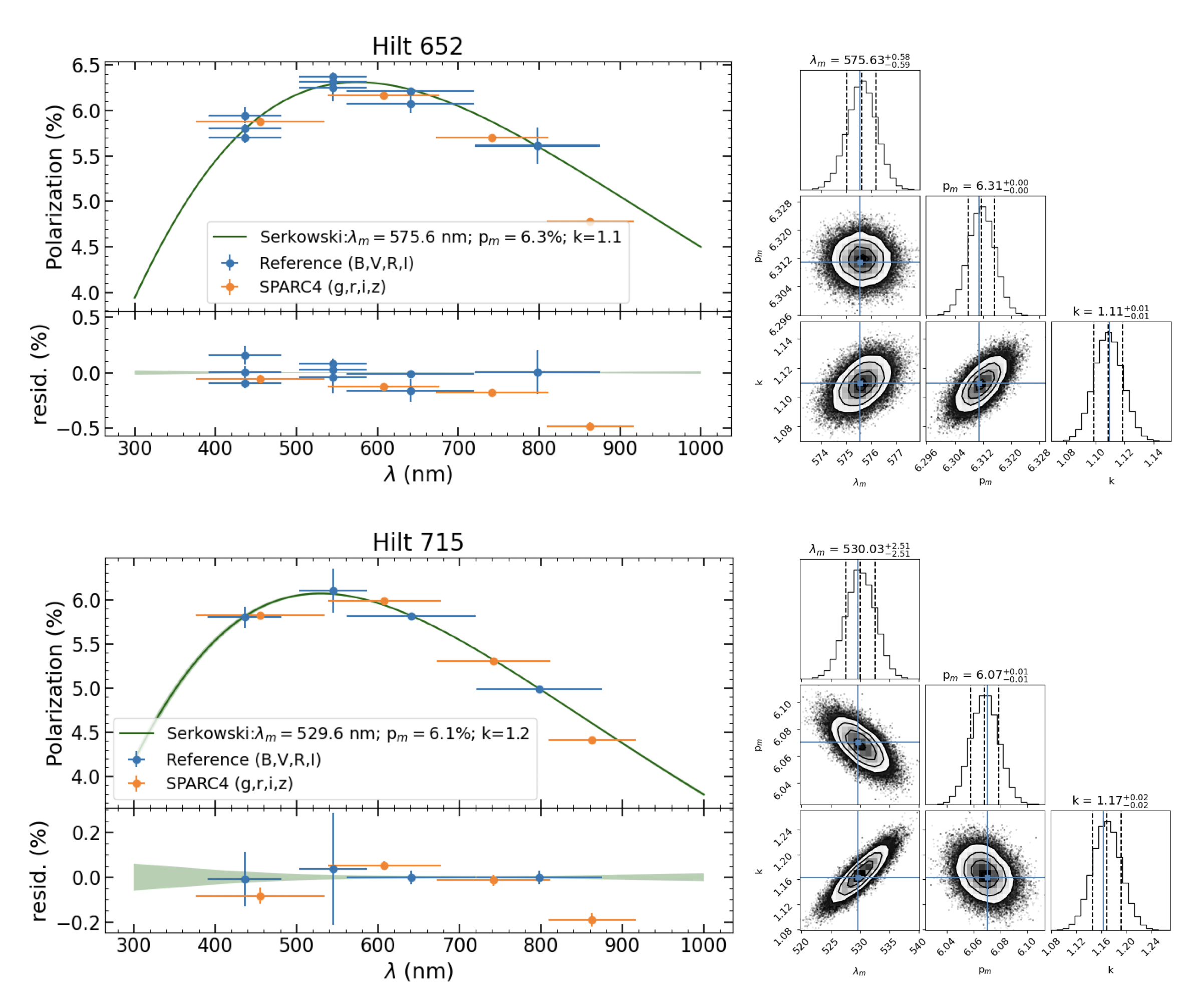}
\caption{Comparison between the linear polarization measured by the SPARC4 Pipeline in the $g$, $r$, $i$, and $z$ bands and literature values in the $B$, $V$, $R$, and $I$ bands for the polarized standard stars Hilt~715 and Hilt~652 \citep{Fossati2007,Cikota2017}. The plots also show the least-squares fits of the Serkowski law to the literature data, with the posterior distributions of the fit parameters displayed in the left panels.}
\label{fig:polar_std_analysis}
\end{figure*}

Reference values from the literature are reported in the Johnson–Morgan photometric system, whereas SPARC4 data are in the SDSS system. We adopt the following pivotal wavelengths and bandwidths for the reference $BVRI$ bands: $\lambda_{B} = 436\pm45$~nm, $\lambda_{V} = 545\pm42$~nm, $\lambda_{R} = 641\pm79$~nm, and $\lambda_{I}=798\pm77$~nm. To compare the SPARC4 with literature values in different bands, we analyzed the polarized standards, Hilt~715 and Hilt~652 as follows. We first least-squares fit the classical Serkowski curve \citep{Serkowski1975} to the literature values of polarization $P_{\lambda}$ as a function of wavelength ($\lambda$):

\begin{equation}
\frac{P(\lambda)}{P_{\rm max}} = \exp{\left[-K\ln^2{\left(\frac{\lambda_{\rm max}}{\lambda}\right)}\right]}, 
\label{eq:serkowski}
\end{equation}

\noindent
where $P_{\rm max}$, $\lambda_{\rm max}$, and $K$ are the maximum polarization, the wavelength of the maximum polarization, and a parameter related to the broadness of the Serkowski curve, respectively.
We sampled the posterior distributions of these parameters using a Bayesian MCMC framework, and the results are shown in Figure~\ref{fig:polar_std_analysis}.  For comparison, we also plotted the SPARC4 polarization measurements, adopting the pivotal wavelengths ($\lambda_{\rm piv}$) for each band from \cite{Bernardes2025a}: $\lambda_{g} = 455\pm79$~nm, $\lambda_{r} = 608\pm69$~nm, $\lambda_{i} = 742\pm70$~nm, and $\lambda_{z} = 863\pm54$~nm.

The differences between the Serkowski fits and the SPARC4 polarization for both standards are within $\Delta P < 0.2\%$ in all bands except the $z$ band. We suspect that the $\lambda_{\rm piv}$ value in the $z$ band could be underestimated, implying that the system throughput is likely greater toward the near-infrared. This will be addressed in an upcoming publication by Bernardes et al. (in prep.). For Hilt~652, the differences are consistently negative, suggesting another systematic effect, possibly related to intrinsic target variability.  

The polarized standard stars are routinely observed to calibrate the equatorial orientation of the instrument’s polarization optics. We calibrated the data for both standards, Hilt~715 and Hilt~652, using the literature values of the polarization angle listed in Table~\ref{tab:polarcalibresults}. The reference polarization angle was taken as the mean of all reference values across all bands, yielding $\bar{\theta}_{\rm ref} = 49.62\pm0.04$~deg for Hilt~715 and $\bar{\theta}_{\rm ref} = 179.42\pm0.03$~deg for Hilt~652. The differences between the measured values and these mean reference values, $\Delta\theta = \theta - \bar{\theta}_{\rm ref}$, are also given in Table~\ref{tab:polarcalibresults}. These $\Delta\theta$ values can be used to calibrate measurements of other targets of unknown polarization observed during the same run. 

The unpolarized star HD~13588 exhibits a polarization of $<0.06$\% with uncertainties of $\sim0.03$\% in all SPARC4 bands. For comparison, \cite{Cikota2017} report polarization values of $<0.12$\% across all bands. These results indicate that the instrumental polarization of SPARC4 is at the level of a few hundredths of a percent or lower.  

\subsection{Time}
\label{sec:time}

The pipeline reads the start time of each exposure, the exposure duration, and the observatory location as provided by the SPARC4 acquisition system in the image headers. It then calculates the mid-exposure time and converts it into several time representations using the \texttt{astropy.time} library. For instance, it computes the barycentric Julian date (BJD) corresponding to the midpoint of exposure for an individual frame and to the midpoint of a sequence of exposures for either a stack or a polarimetric measurement. The BJD is the reference time used by the pipeline for constructing both photometric and polarimetric time series.

%--------------------------------------------------------------------
\section{Analysis of exoplanet transit light curves}
\label{sec:transit-analysis}

The analysis of the time series data for the transits of exoplanets observed with SPARC4 and TESS/K2 (Tables \ref{tab:sparc4observations} and \ref{tab:tessobservations}) was conducted following the same approach as in \cite{Martioli2022,Martioli2023,Martioli2024}, summarized as follows. To focus on the data most relevant for transit fitting and to allow efficient modeling of the baseline flux, we retained only the TESS/K2 observations within $T_{c} \pm 1.25 \times t_{d}$, where $T_{c}$ is the time of conjunction (mid-transit), and $t_{d}$ is the transit duration. Figure~\ref{fig:tess_k2_transits} illustrates the detrended TESS and K2 data adopted in this analysis. Our framework uses a transit model implemented in the \texttt{BATMAN} package \citep{Kreidberg2015}, assuming a quadratic limb-darkening law that is initially taken to be constant across all spectral bands. For simplicity, we assumed circular orbits for all seven exoplanets, fixing the orbital eccentricity to zero and the argument of periastron to 90 degrees. 

The initial planetary parameters were adopted from published values, which are summarized in Table \ref{tab:planetparams}. We first applied a least-squares fit using the \texttt{scipy.optimize.leastsq} tool with simultaneous modeling of a detrending polynomial baseline and the transit model. This was followed by a Bayesian MCMC analysis with the \texttt{emcee} package \citep{foreman2013} to derive the posterior probability distributions of the system parameters. For each analysis, we used 50 walkers over 1,500 iterations, discarding the first 500 samples as burn-in.

\begin{table*}
\centering
\tiny
\caption{Posterior distributions of the system parameters derived from transit modeling. For each object, we present the literature reference values, the fits using TESS (or K2) data alone, the joint fits combining TESS (or K2) and SPARC4 data, and the results from independent fits of the transits observed in each SPARC4 band.}
\label{tab:planetparams} 
\begin{tabular}{cccccccccc}
\hline\hline
\multirow{3}{*}{Object ID} & Time of & Orbital &  Scaled semi- & Planet-to-star &  Orbital & \multicolumn{2}{c}{Limb darkening} & Error & \multirow{3}{*}{Reference} \\
& conjunction & period &  major axis & radius ratio & inclination & \multicolumn{2}{c}{coefficients} & benchmark & \\
          & T$_{c}$ (BJD-2400000) & $P_{\rm orb}$ (d)      &    $a/R_{\star}$ &       $R_{p}/R_{\star}$     & $i$ (deg)  & $u_{0}$  &    $u_{1}$  & $\langle \sigma_{\rm stddev}/\bar{\sigma}\rangle $ &  \\
\hline
\multirow{7}{*}{WASP-78}  & $55882.3588(5)$ & $2.1751763(5)$ & $3.5\pm1.6$ & $0.0794(13)$ & $83.2^{+2.3}_{-1.6}$ & 0.388 & 0.609 && \cite{Smalley2012} \\                  
                          & $58438.2016(5)$ & $2.1751843(16)$ & $3.82\pm0.05$ & $0.0850(5)$ & $88.1\pm1.6$  & $0.27(8)$ & $0.15(13)$ && only TESS data  \\
                          & $\textbf{58438.20150(35)}$ & $\textbf{2.1751844(5)}$ & ${\bf 3.83\pm0.05}$ & $\textbf{0.08573(29)}$ & $\bf{87.4\pm1.3}$ & $0.27(30)$ & $0.09(6)$ && \textbf{TESS+SPARC4} \\ 
                          & $58438.20151(23)$ & FIXED & $3.67\pm0.15$ & $0.0870(11)$ & $82.8\pm2.3$ & $0.02(7)$ & $0.42(19)$ &$0.7\pm0.1$&  SPARC4 g band   \\              
                          & $58438.20137(12)$ & FIXED & $3.825\pm0.021$ & $0.08481(26)$ & $89.0\pm1.2$  & $0.07(4)$ & $0.55(9)$ & $1.1\pm0.1$&  SPARC4 r band  \\            
                          & $58438.20145(15)$ & FIXED & $3.870\pm0.010$ & $0.08517(24)$ & $89.9\pm0.8$  & $0.35(2)$ & $0.004(21)$ &$1.1\pm0.1$&  SPARC4 i band   \\         
                          & $58438.20168(23)$ & FIXED & $3.890\pm0.018$ & $0.08651(35)$ & $89.6\pm1.3$  & $0.20(6)$ & $0.04(10)$ &$1.1\pm0.1$&  SPARC4 z band  \\             
 \hline
 \multirow{7}{*}{HATS-23} & $57072.8527(7)$ & $2.1605160(50)$ & $6.1^{0.4}_{-0.3}$ & $0.159(20)$ & $81.0^{+0.9}_{-0.6}$ & 0.3728 & 0.3215 && \cite{Bento2017} \\             
                            & $58654.3462(11)$ & $2.1605122(21)$ & $8.7\pm1.8$ & $0.099(16)$ & $86.3\pm2.9$ & $0.8(4)$ & $0.4(4)$ && only TESS data   \\           
                            & $\textbf{58654.3463(11)}$ & $\textbf{2.1605120(13)}$ & $\bf{7.20\pm0.18}$ & $\textbf{0.1145(15)}$ & $\bf{83.92\pm0.27}$  & $0.70(25)$ & $0.50(31)$ && \textbf{TESS+SPARC4 } \\  
                            & $58654.34689(18)$ & FIXED & $7.16\pm0.15$ & $0.1133(26)$ & $83.71\pm0.22$  & $0.26(28)$ & $0.8(4)$ &$1.0\pm0.2$& SPARC4 g band  \\             
                            & $58654.34618(8)$ & FIXED & $6.95\pm0.17$ & $0.1143(16)$ & $83.81\pm0.24$   & $0.97(23)$ & $0.30(27)$ &$1.0\pm0.1$& SPARC4 r band  \\
                            & $58654.34652(11)$ & FIXED & $7.66\pm0.17$ & $0.1097(23)$ & $84.57\pm0.24$  & $0.54(27)$ & $0.52(35)$ &$1.1\pm0.1$& SPARC4 i band  \\            
                            & $58654.34529(18)$ & FIXED & $7.48\pm0.19$ & $0.1102(27)$ & $84.30\pm0.27$  & $0.48(32)$ & $0.64(40)$ &$1.2\pm0.1$& SPARC4 z band  \\            
 \hline                  
\multirow{7}{*}{HATS-24}  & $57038.4733(4)$ & $1.3484954(13)$ & $4.67^{+0.10}_{-0.14}$ & $0.131(3)$ & $86.6\pm1.2$ & 0.2638 & 0.3753 &&  \cite{Bento2017}\\                  
                            &  $58653.97351(19)$ & $1.34849651(27)$ & $4.41\pm0.09$ & $0.1317(8)$ & $84.2\pm0.6$ & $0.27(7)$ & $0.10(12)$ && only TESS data   \\   
                            &  $\textbf{58653.97351(28)}$ & $\textbf{1.34849651(27)}$ & $\bf{4.40\pm0.04}$ & $\textbf{0.1308(5)}$ & $\bf{84.17\pm0.21}$ & $0.41(8)$ & $0.034(35)$ && \textbf{TESS+SPARC4}  \\  
                            &  $58653.9714(6)$ & FIXED & $4.42\pm0.05$ & $0.1291(12)$ & $83.53\pm0.28$   & $0.03(9)$ & $0.86(20)$ &$0.8\pm0.1$& SPARC4 g band  \\
                            &  $58653.9737(4)$ & FIXED & $4.348\pm0.031$ & $0.1300(4)$ & $84.11\pm0.21$   & $0.465(36)$ & $0.03(6)$ &$1.0\pm0.1$& SPARC4 r band  \\
                            &  $58653.9755(6)$ & FIXED & $4.35\pm0.05$ & $0.1312(4)$ & $84.79\pm0.29$   & $0.333(28)$ & $0.01(4)$ &$1.0\pm0.1$& SPARC4 i band  \\             
                            &  $58653.9742(5)$ & FIXED & $4.62\pm0.07$ & $0.1279(6)$ & $86.9\pm0.6$   & $0.454(28)$ & $0.008(38)$ &$1.0\pm0.2$& SPARC4 z band  \\
 \hline                   
 \multirow{7}{*}{HATS-9}  & $56124.2590(9)$ & $1.915307(5)$ & $4.36^{+0.10}_{-0.25}$ & $0.073(4)$ & $86.5^{+1.6}_{-2.5}$ & 0.4688 & 0.2596 &&  \cite{Brahm2015} \\                   
                            & $57302.17401(7)$ & $1.9153344(31)$ & $3.241\pm0.014$ & $0.0893(5)$ & $77.40\pm0.14$ & $0.24(9)$ & $0.56(13)$ && only K2 data   \\ 
                            & $\textbf{57302.17449(5)}$ & $\textbf{1.91531219(4)}$ & $\bf{3.66\pm0.06}$ & $\textbf{0.08780(30)}$ & $\bf{80.5\pm0.4}$ & $0.58(4)$ & $0.07(6)$ && \textbf{K2+SPARC4}  \\  
                            & $57302.17383(11)$ & FIXED & $4.13\pm0.05$ & $0.0836(6)$ & $84.5\pm0.5$   & $0.788(26)$ & $0.01(4)$ &$0.8\pm0.1$& SPARC4 g band  \\
                            & $57302.17329(4)$ & FIXED & $4.361\pm0.033$ & $0.08414(22)$ & $86.1\pm0.4$  & $0.570(9)$ & $0.003(13)$ &$0.9\pm0.1$& SPARC4 r band  \\   
                            & $57302.17335(5)$ & FIXED & $4.473\pm0.034$ & $0.08381(21)$ & $87.6\pm0.6$  & $0.501(12)$ & $0.007(16)$ &$0.9\pm0.1$& SPARC4 i band  \\ 
                            & $57302.17306(9)$ & FIXED & $4.11\pm0.05$ & $0.08611(31)$ & $84.1\pm0.4$  & $0.443(19)$ & $0.008(19)$ &$1.0\pm0.1$& SPARC4 z band  \\      
 \hline                 
 \multirow{7}{*}{WASP-123} & $56845.1708(4)$ & $2.9776412(23)$ & $7.13\pm0.25$ & $0.1054(13)$ & $85.7\pm0.6$ & $0.683$ & $-0.405$ && \cite{Turner2016} \\
                             & $58655.57897(14)$ & $2.9776446(5)$ & $7.23\pm0.16$ & $0.1036(12)$ & $86.2\pm0.4$  & $0.18(10)$ & $0.54(20)$ &&  only TESS data   \\  
                             & $\textbf{58655.57898(15)}$ & $\textbf{2.97764503(33)}$ & $\bf{7.46\pm0.07}$ & $\textbf{0.1036(4)}$ & $\bf{86.68\pm0.18}$  & $0.38(4)$ & $0.23(8)$ &&  \textbf{TESS+SPARC4}  \\        
                             & $58655.5790(15)$ & FIXED & $7.29\pm0.24$ & $0.1021(19)$ & $86.1\pm0.5$ & $0.31(17)$ & $0.67(35)$  &$0.7\pm0.1$&  SPARC4 g band   \\                   
                             & $58655.5793(7)$ & FIXED & $7.53\pm0.08$ & $0.1041(5)$ & $86.71\pm0.19$  & $0.34(6)$ & $0.24(12)$ &$1.0\pm0.1$&  SPARC4 r band  \\                   
                             & $58655.5786(9)$ & FIXED & $7.58\pm0.10$ & $0.1036(5)$ & $87.14\pm0.25$ & $0.43(6)$ & $0.15(12)$ &$0.9\pm0.1$&  SPARC4 i band   \\                   
                             & $58655.5786(8)$ & FIXED & $7.42\pm0.09$ & $0.1039(5)$ & $86.81\pm0.22$  & $0.49(4)$ & $0.03(7)$ &$0.9\pm0.2$&  SPARC4 z band  \\     
 \hline                   
 \multirow{7}{*}{HATS-21}  & $57109.2254(6)$ & $3.554397(6)$ & $9.8^{+0.4}_{-0.8}$ & $0.113(11)$ & $85.0^{+0.2}_{-0.7}$ & 0.3952 & 0.3072 &&  \cite{Bhatti2016} \\                   
                            & $58655.3935(4)$ & $3.5544073(16)$ & $9.3\pm0.5$ & $0.1064(34)$ & $84.9\pm0.4$ & $0.42(36)$ & $0.6(5)$ && only TESS data   \\         
                            &  $\textbf{58655.3940(5)}$ & $\textbf{3.5544053(12)}$ & $\bf{9.14\pm0.11}$ & $\textbf{0.1095(17)}$ & $\bf{84.55\pm0.10}$  & $0.33(22)$ & $0.36(25)$ && \textbf{TESS+SPARC4}  \\                       
                            &  $58655.3913(28)$ & FIXED & $9.5\pm0.4$ & $0.105(4)$ & $84.72\pm0.18$  & $0.04(13)$ & $0.07(17)$ &$0.8\pm0.1$& SPARC4 g band  \\
                            &  $58655.3922(6)$ & FIXED & $9.29\pm0.15$ & $0.1139(27)$ & $84.72\pm0.10$  & $0.64(32)$ & $0.2(4)$ &$0.9\pm0.1$& SPARC4 r band  \\             
                            &  $58655.3926(8)$ & FIXED & $9.44\pm0.17$ & $0.1140(30)$ & $84.79\pm0.13$ & $0.08(26)$ & $0.63(34)$ &$0.9\pm0.1$& SPARC4 i band  \\             
                            &   $58655.3905(6)$ & FIXED & $9.51\pm0.18$ & $0.114(6)$ & $84.54\pm0.18$  & $0.11(24)$ & $0.07(35)$ &$0.8\pm0.1$& SPARC4 z band  \\       
 \hline                  
 \multirow{7}{*}{WASP-111} & $56275.7513(4)$ & $2.3109650(24)$ & $4.57\pm0.20$ & $0.0802(12)$ & $81.6\pm0.8$ & 0.488 & 0.454 && \cite{Anderson2014} \\           
                             & $58325.58231(16)$ & $2.31097030(31)$ & $4.42\pm0.07$ & $0.0816(4)$ & $81.11\pm0.31$  & $0.21(12)$ & $0.22(15)$ && only TESS data   \\ 
                             & $\textbf{58325.58207(16)}$ & $\textbf{2.31097137(18)}$ & $\bf{4.58\pm0.04}$ & $\textbf{0.08047(29)}$ & $\bf{82.00\pm0.20}$ & $0.08(6)$ & $0.44(9)$ && \textbf{TESS+SPARC4}  \\ 
                             & $58325.58326(15)$ & FIXED & $4.70\pm0.07$ & $0.0786(5)$ & $82.75\pm0.33$  & $0.02(5)$ & $0.77(9)$ &$0.6\pm0.1$&  SPARC4 g band   \\                  
                             & $58325.58255(7)$ & FIXED & $4.505\pm0.031$ & $0.08048(21)$ & $81.69\pm0.15$  & $0.03(6)$ & $0.44(8)$ &$0.8\pm0.1$&  SPARC4 r band  \\                
                             & $58325.58188(9)$ & FIXED & $4.55\pm0.04$ & $0.08131(35)$ & $81.80\pm0.20$  & $0.03(10)$ & $0.32(14)$ &$0.7\pm0.1$&  SPARC4 i band   \\     
                             & $58325.58166(12)$ & FIXED & $4.62\pm0.04$ & $0.08127(28)$ & $82.18\pm0.20$  & $0.39(7)$ & $0.02(9)$ &$0.6\pm0.1$&  SPARC4 z band  \\                                   
 \hline                   
\hline
\end{tabular}
\end{table*} 

To benchmark our results obtained with SPARC4 data, we first fitted only the TESS data (or K2 data in the case of HATS-9) to obtain best-fit parameters for each system. Uniform prior distributions were adopted with the following bounds. For the time of conjunction, $T_{c}$, we considered bounds within $\pm10\%$ of the orbital period around the central time of the first epoch observed. For the orbital period, $P$, within $\pm20\%$ of the literature values. All other parameters were allowed to vary within bounds extending at least $\pm10\sigma$ from their literature values, or up to their physical limits. Table \ref{tab:planetparams} presents the results of this analysis in the rows labeled as ``TESS (or K2) data''. The TESS and K2 light curves and their corresponding best-fit models are shown in Appendix~\ref{app:s4transitfit}.
 
\begin{figure*}
\centering
\includegraphics[width=1.\hsize]{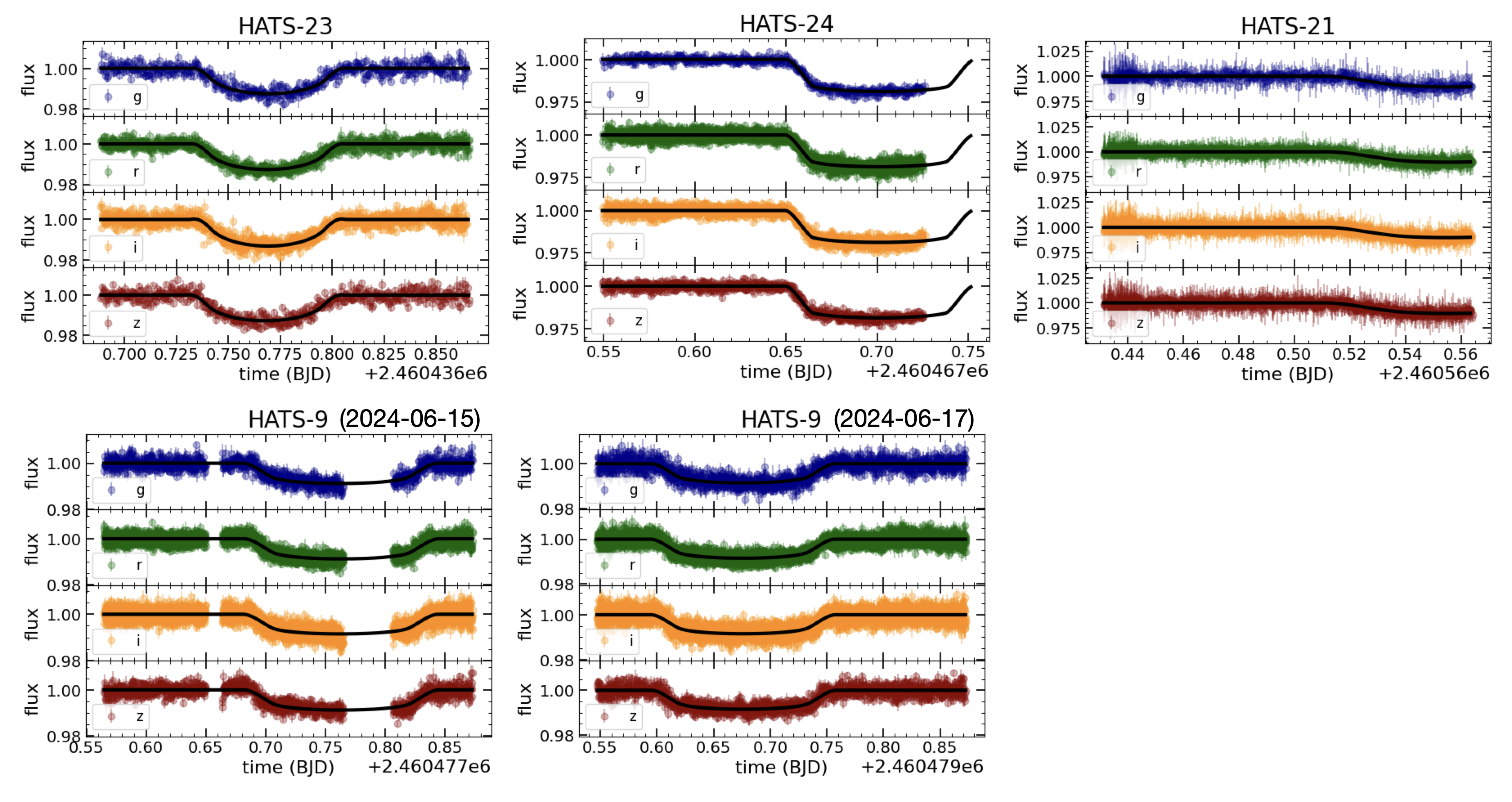}
\caption{Four-band SPARC4 differential photometry time series of the transits of the exoplanets HATS-23~b, HATS-24~b, HATS-21~b, and HATS-9~b. Each panel show the observed light curve data with best-fit transit models (black lines).}
\label{fig:phot_transits}
\end{figure*}

The SPARC4 photometric data consist of one differential light curve per comparison star, each treated as an independent dataset with its baseline calibration fitted separately, as in \cite{Martioli2018}. We then performed a joint analysis combining the TESS/K2 data around all transits with the SPARC4 light curves from all comparison stars and bands, adopting the previously derived parameters as initial values and maintaining uniform priors. Figure~\ref{fig:phot_transits} illustrates the final SPARC4 light curve data and the best-fit transit models. Table \ref{tab:planetparams} presents the results of this analysis in the rows labeled as ``TESS (or K2)+SPARC4''. This joint solution provides the best constraints on the planetary parameters and is therefore adopted as our final result (highlighted in boldface in Table \ref{tab:planetparams}).  However, it does not constrain the limb-darkening coefficients (LDC), as these are expected to vary across different spectral bands \citep{Claret2011,Claret2017}. 

Thus, we fitted the SPARC4 light curves for each photometric band independently as illustrated in Appendix~\ref{app:s4transitfit}, adopting uniform priors for all parameters except the orbital period, which was fixed since it cannot be constrained from a single transit. The fitted transit times have uncertainties on the order of minutes or less, indicating that our data provide consistent timing and have sufficient precision to search for transit-timing variations (TTVs). This analysis also allows us to assess the spectral dependence of both the limb-darkening coefficients and the planet-to-star radius ratio. However, caution is required when interpreting these results, as some SPARC4 observations covered only partial transits, which limits the accuracy of the derived parameters. The fitted quadratic LDCs have uncertainties on the order of $1$–$10\%$, which can be compared with theoretical predictions \citep{Morello2017}. Obtaining simultaneous multi-band LDCs is particularly important for accurately determining the planet-to-star radius ratio, $R_p/R_\star$. Our per-band fitted values of $R_p/R_\star$ already reach a precision of $\sim10^{-4}$, even with unconstrained LDCs.

\begin{figure*}
\centering
\includegraphics[width=1.\hsize]{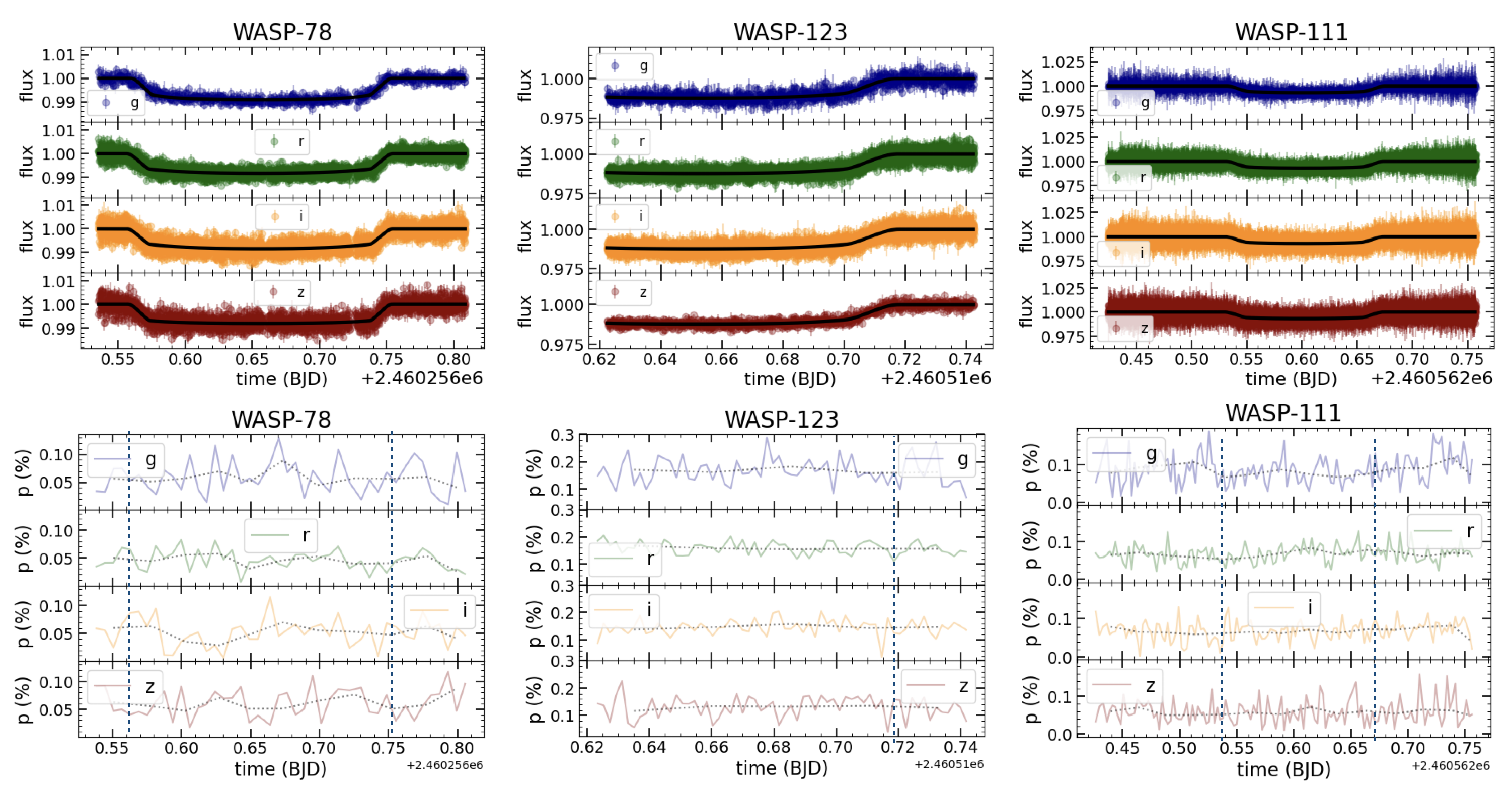}
\caption{Four-band SPARC4 differential photometry (top panels) and polarimetric time series (bottom panels) for the transits of the exoplanets WASP-78 b (left), WASP-123 b (middle), and WASP-111 b (right). The top panels present the observed light curves together with the best-fit transit models (black lines). The bottom panels show the corresponding total polarization using the same color scheme as in the photometric plots. Dotted lines indicate the data binned in 30-minute intervals, and vertical dashed lines mark the times of transit ingress and egress. }
\label{fig:polar_transits}
\end{figure*}

We evaluate the photometric precision and consistency of the pipeline uncertainties using the residual light curves after subtracting the best-fit transit models from the data shown in Figures~\ref{fig:phot_transits} and \ref{fig:polar_transits}. To benchmark the pipeline uncertainties, we computed the ratio between the dispersion of residuals, $\sigma_{\rm stddev}$ (defined as the standard deviation within 15-minute bins), and the median per-point uncertainty, $\bar{\sigma}$, evaluated over the same bins. We then calculated the mean and standard deviation of this ratio over the full time series. The results for each SPARC4 band are reported in the ``Error benchmark'' column of Table~\ref{tab:planetparams}, yielding values close to unit, indicating residual dispersions comparable to the pipeline uncertainties.

\begin{figure*}
\centering
\includegraphics[width=1.\hsize]{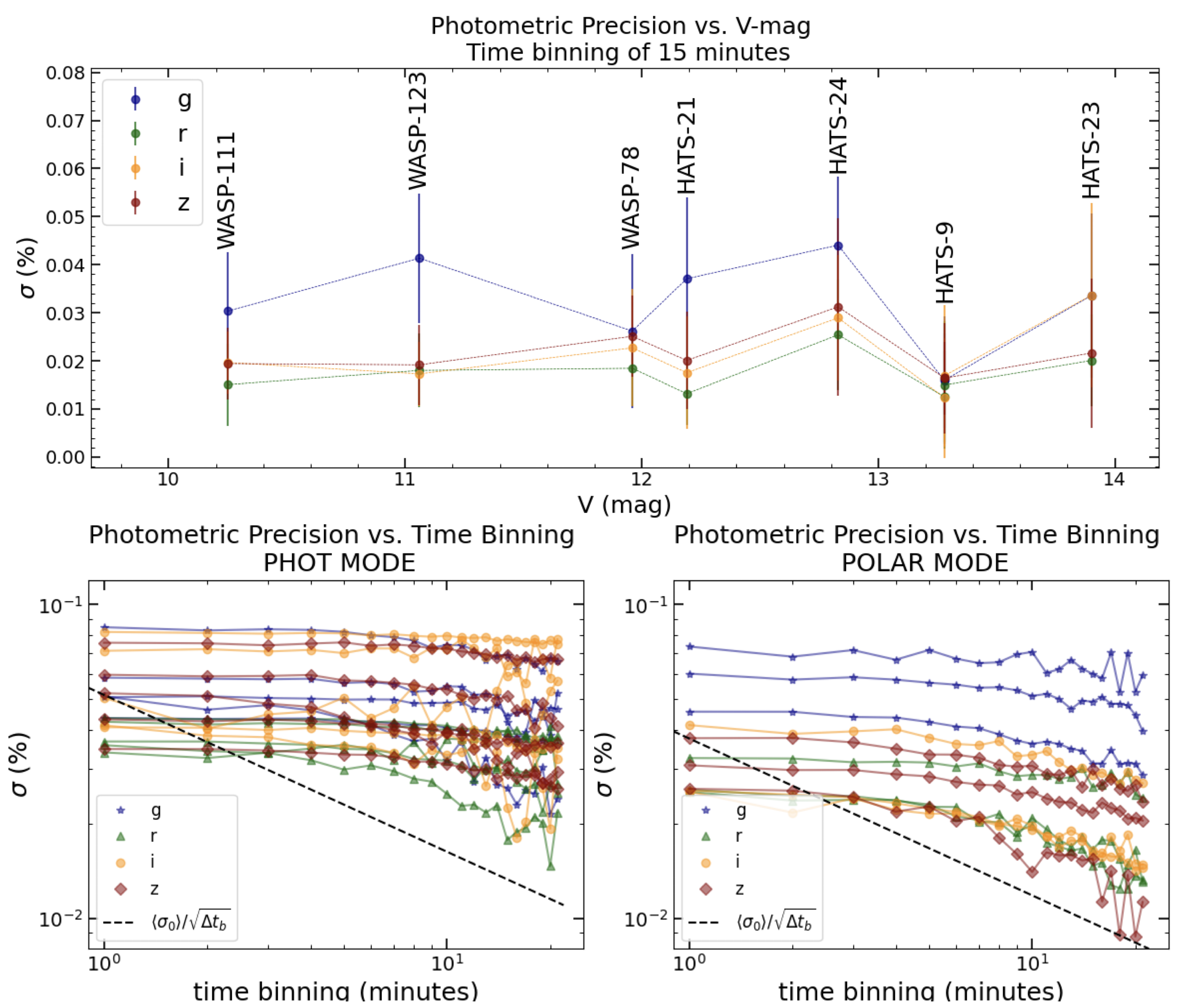}
\caption{Photometric precision of SPARC4 differential light-curve residuals. The top panel shows the precision as a function of target $V$ magnitude, computed from the mean standard deviation of residuals in 15-minute bins. The bottom panels show the precision as a function of bin size for all targets and channels in PHOT (left) and POLAR (right) modes. Dashed lines show the expected binning improvement, $\langle \sigma_{0} \rangle / \sqrt{\Delta t}$, where $\langle \sigma_{0} \rangle$ represents the average precision without binning and $\Delta t_b$ the bin size. The achieved precision spans 0.01\%–0.08\%.}
\label{fig:timeseries_phot_precision}
\end{figure*}

Figure~\ref{fig:timeseries_phot_precision} summarizes final photometric precision achieved in our analysis. In the top panel, we show the precision as a function of target $V$ magnitude for the four SPARC4 channels. Precision is computed as the mean standard deviation of the residuals in 15-minute bins, ranging from 0.01\% to 0.04\%, with no strong dependence on target magnitude. This is because the photometric precision also depends on the properties of the comparison stars, which are not represented in this plot. The average photometric precision for all targets in all bands is $0.022 \pm 0.002$\%. To account for cadence, the bottom panels of Figure~\ref{fig:timeseries_phot_precision} show the photometric precision as a function of bin size for all targets and channels. The precision, estimated as the standard deviation of the residuals, ranges from 0.01\% to 0.08\%. As expected, increasing the bin size improves precision at the expense of temporal resolution.

%\subsection{Timing accuracy based on mid-transit}

%To verify the timing accuracy—both from the SPARC4 acquisition system and from the pipeline's calculations of BJD—we compared the expected times of mid-transit ($T_c$), derived from ephemerides based on transit model fits to photometric time series from TESS and K2, with the best-fit time of mid-transit obtained in each SPARC4 channel independently, as presented in Table \ref{tab:planetparams}. These results are illustrated in Figure \ref{fig:timeaccuracy}, where one can see that there is no obvious time offset.  The standard deviation has been estimated in 85 seconds, which is much larger than the time accuracy of the SPARC4 acquisition system, estimated to be on the order of miliseconds.  The measured mid-transit times can have additional uncertainties from the fact that not all transits are fully observed with proper out-of-transit baselines or due to uncertainties in the other parameters. 

%\begin{figure}
%\centering
%\includegraphics[width=1.\hsize]{Figures/s4_time_accuracy.png}
%\caption{Transit time difference, $\Delta T_{c}$, defined as the difference between the times of mid-transit $T_c$ fitted in each SPARC4 channel and the $T_c$ fitted using only TESS/K2 data. The dotted grey lines indicate the standard deviation of all measurements, estimated as 85 seconds. }
%\label{fig:timeaccuracy}
%\end{figure}

%--------------------------------------------------------------------
\section{Analysis of polarimetric time series}
\label{sec:polar-analysis}

The SPARC4 observations of WASP-78, WASP-123, and WASP-111 were carried out in POLAR L2 mode, which allowed us to obtain simultaneous photometric and polarimetric time series for these targets. One of the goals of this experiment was to test whether the use of polarimetric optics and the subsequent data reduction would affect photometric precision. Moreover, during exoplanet transits, a polarimetric signal is expected due to the breaking of spherical symmetry when the planet passes in front of the stellar disk \citep{Carciofi2005, Kostogryz2015}. This polarization signal is predicted to be on the order of $10^{-4}$~\% for a Jupiter-sized planet transiting a Sun-like star, and, in the most favorable cases, up to $10^{-1}$~\% for a planet with a radius of $2\,R_{\rm Jup}$ transiting an M dwarf. Therefore, we also estimate the polarimetric precision in our hours-long time series.

\begin{figure}[ptbh]
\centering
\includegraphics[width=0.45\textwidth]{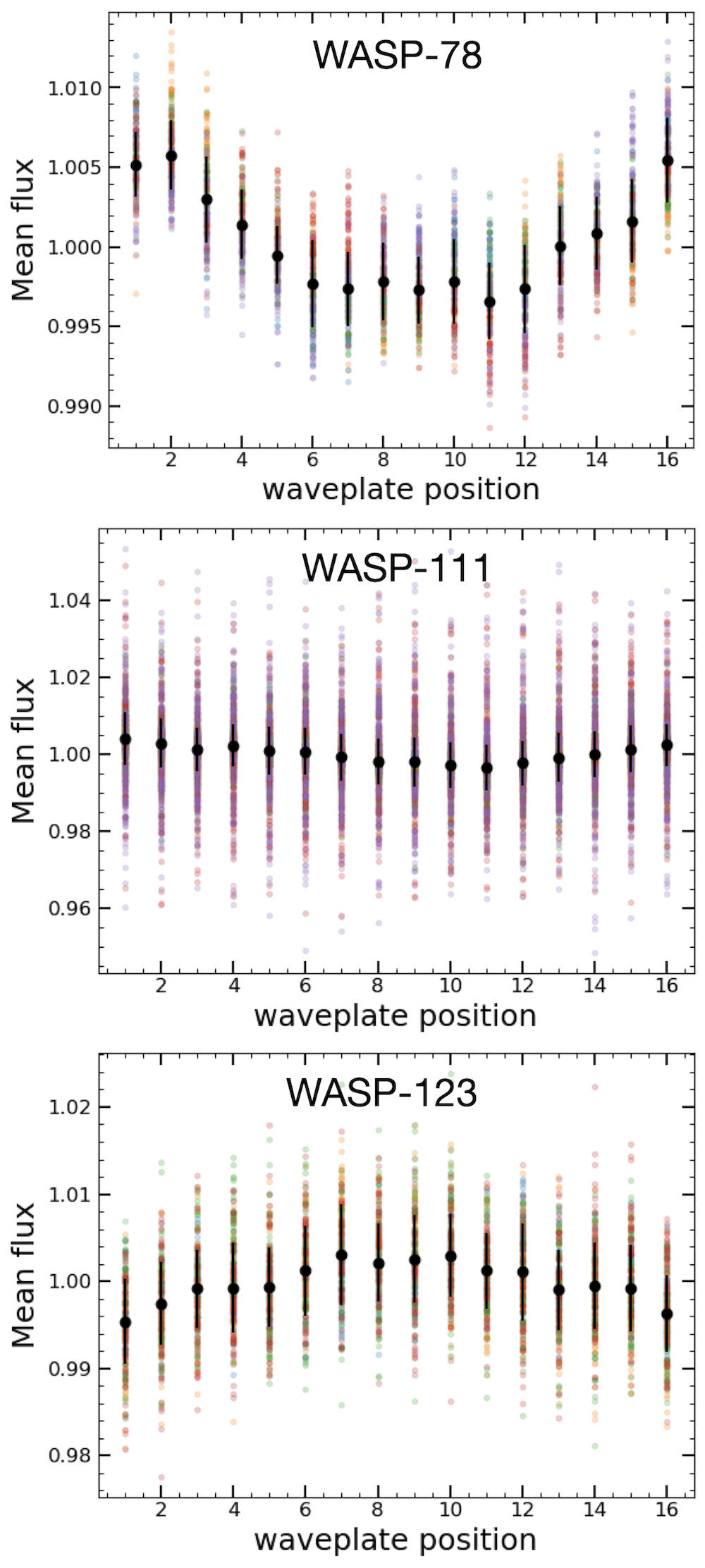}
\caption{Flux modulation as a function of waveplate position for the $g$-band SPARC4 data from three transits observed in POLAR L2 mode. Points of the same color correspond to the same 16-position polarimetric sequence, while black points with error bars indicate the median and median absolute deviation at each waveplate position.}
\label{fig:s4_wppos}
\end{figure}

\begin{figure}[ptbh]
\centering
\includegraphics[width=0.45\textwidth]{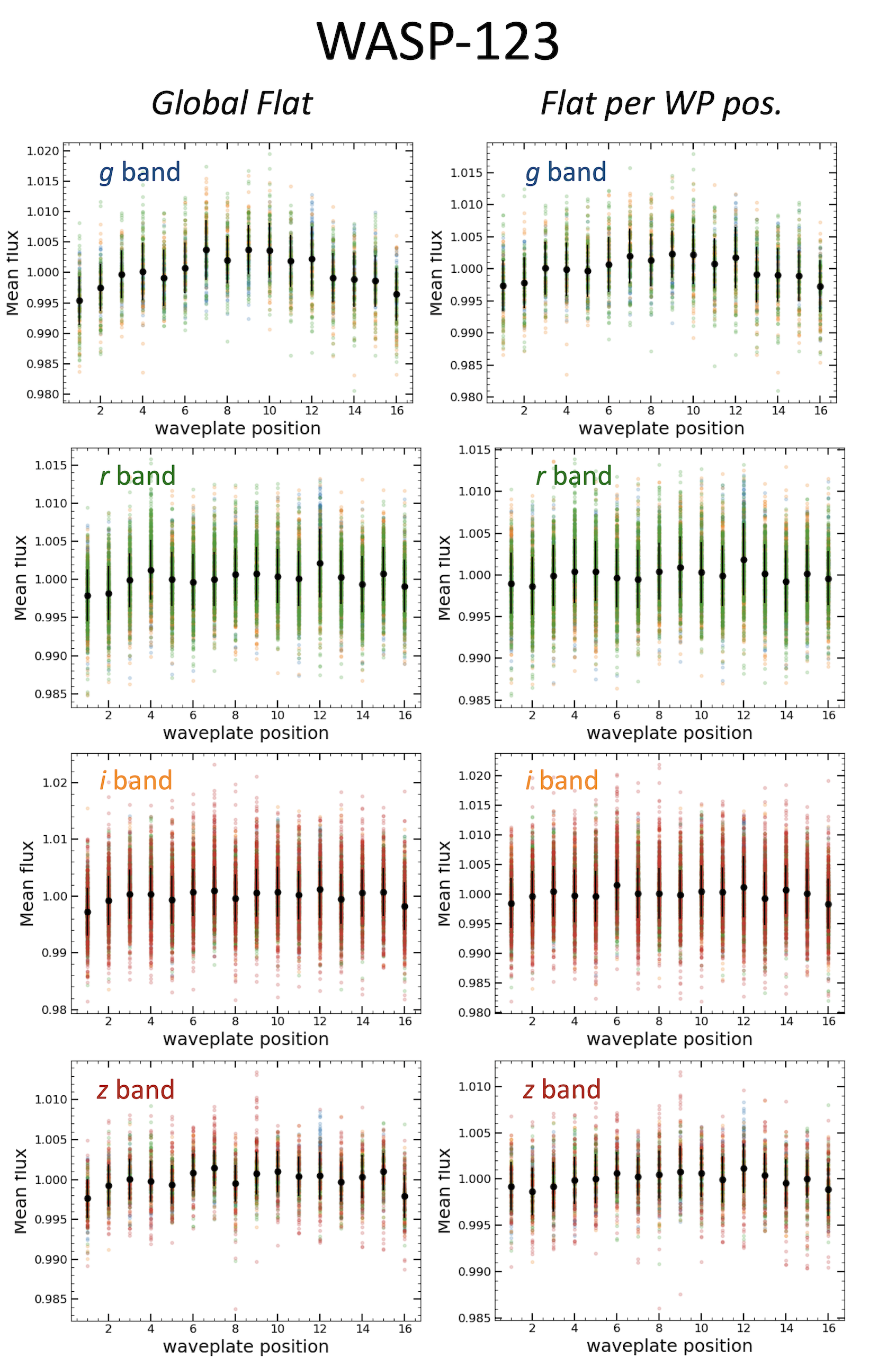}
\caption{Flux modulation as a function of waveplate position for the four-band SPARC4 observations of the WASP-123 b transit in POLAR L2 mode. Points of the same color correspond to the same 16-position polarimetric sequence; black points with error bars show the median and median absolute deviation at each position. Left panels use a global flat-field (combining flats from all waveplate positions), while right panels use position-dependent flat-fields.}
\label{fig:s4_wppos_wasp123}
\end{figure}

We found that photometric precision is affected by the rotation of the waveplate, as illustrated in Figures~\ref{fig:s4_wppos} and \ref{fig:s4_wppos_wasp123}. In Figure~\ref{fig:s4_wppos} we show the SPARC4 differential photometry data in the $g$ band after removing the transit signal of the planet as a function of the waveplate rotation position (WPPOS) from 1 to 16. There is a clear modulation of the differential flux with the waveplate position with an amplitude up to $\sim1\%$, which is small but important for the level required to detect planetary transits. 

As of now, the pipeline does not correct for this modulation automatically as it appears to have a variable pattern. In our analysis presented here, we measured and removed this modulation from all observations in POLAR mode. However, the amplitude of this modulation decreases rapidly to redder bands, making it particularly more important for the g band (see Figure~\ref{fig:s4_wppos_wasp123}). 

A way to mitigate this problem is to obtain flat-field calibration data for each position of the waveplate. The pipeline detects automatically the calibration data obtained in each position and builds a master flat per position and applies the flat-field correction following the same position. 

Figure~\ref{fig:s4_wppos_wasp123} (right panel) shows the effect of this correction in the WASP-123 data. Table \ref{tab:wpposeffect_wasp123} reports the flux dispersion before and after applying the correction, obtained from the mean modulation after removing the transit signal. This correction improves the photometric precision in the $g$ band by 0.14\% when using a global flat, and by 0.08\% when using flats per waveplate position. In the other bands, the gain becomes negligible when flats per waveplate position are applied. Therefore, for other scientific applications where the signal is not known, flats should always be obtained for each waveplate position to ensure the best possible photometric precision.

%\begin{deluxetable*}{cccccccc}
%\tablecaption{Final dispersion of the SPARC4 light curve of WASP-111, estimated using the median absolute deviation, before and after WPPOS detrending. Results are shown for reductions with a global flat (combined from all waveplate positions) and with flats calculated separately for each waveplate position.}
%\label{tab:wpposeffect_wasp111}
%\tablehead{\colhead{Flat-field} & \colhead{Detrended} & \colhead{$g$ band} & \colhead{$r$ band} & \colhead{$i$ band} & \colhead{$z$ band} }
%\startdata
%global & No & 0.72\% & 0.59\% & 0.73\% & 0.78\% \\
%global & Yes & 0.57\% & 0.57\% & 0.72\% & 0.77\% \\
%per WPPOS & No & 0.64\% & 0.58\% & 0.73\% & 0.77\% \\
%per WPPOS & Yes & 0.56\% & 0.56\% & 0.72\% & 0.76\% \\
%\enddata
%\end{deluxetable*} 

\begin{deluxetable}{cccccccc}
\tablecaption{Final dispersion of the SPARC4 light curve of WASP-123, estimated using the median absolute deviation, before and after WPPOS detrending. Results are shown for reductions with a global flat (combined from all waveplate positions) and with flats calculated separately for each waveplate position.}
\label{tab:wpposeffect_wasp123}
\tablehead{\colhead{Flat-field} & \colhead{Detrended} & \colhead{$g$ band} & \colhead{$r$ band} & \colhead{$i$ band} & \colhead{$z$ band} }
\startdata
global & No & 0.50\% & 0.37\% & 0.44\% & 0.28\% \\
global & Yes & 0.43\% & 0.36\% & 0.43\% & 0.25\% \\
per WPPOS & No & 0.45\% & 0.37\% & 0.44\% & 0.26\% \\
per WPPOS & Yes & 0.41\% & 0.36\% & 0.43\% & 0.26\% \\
\enddata
\end{deluxetable} 

Finally, we analyzed the polarization time series to search for a possible polarimetric signal during the transits. As shown in the bottom panels of Figure~\ref{fig:polar_transits}, none of the three transits exhibited any clear feature in the polarimetric data.  In the best cases, we achieved a polarization dispersion of $\sigma_{\rm p} \sim 0.02\%$ in the $r$ band, which is considered excellent, yet still insufficient to detect the expected signal for these targets. 

%-----------------------------------------------------------------
\section{Conclusions}
\label{sec:conclusions}

We developed a new astronomical data-reduction package in Python, implemented as a pipeline in which a single command performs the complete reduction of SPARC4 data acquired in one night. The pipeline includes basic calibration, astrometry, aperture photometry, and polarimetry. The results are saved in FITS files. We used SPARC4 observations of eight transits of seven hot Jupiters - five obtained in photometric mode and three in polarimetric mode - to perform a scientific validation of the pipeline and to assess the quality of SPARC4 data. 

Astrometric solutions from the pipeline reach sub-arcsecond accuracy in stacked images, even for sparse fields. Using nightly calibrations based on the SkyMapper catalog, we achieved an absolute photometric precision of $\sim0.13$~mag. Differential photometry of the time series yields a typical relative precision of $\sim0.02$\% for stars with V magnitudes between 10 and 14, at a 15-minute cadence. Observations of polarimetric standard stars show that pipeline can achieve a polarimetric accuracy of the order of $0.2\%$, which can likely be improved with refined calibration. Observations of a non-polarized standard star indicate an instrumental polarization of $<0.06$\% in all SPARC4 bands. Our measurements impose an upper limit for the linear polarization in the exoplanet light curves of $\sigma_{\rm p} \sim 0.02\%$ in the r band. This limit was estimated by the dispersion of the linear polarization measurements.

We analyzed the light-curve data of seven planetary systems and obtained planetary parameters that are consistent with previously reported values, while in most cases providing tighter constraints on their orbital and physical properties. This improvement results from combining all available TESS/K2 data with extended temporal coverage from SPARC4 transits, and from the high precision of SPARC4 multi-band photometry processed with a systematic reduction pipeline. Together, these factors allow more accurate determinations of the orbital parameters, the planet-to-star radius ratio, and color-dependent limb-darkening coefficients.  In summary, our homogeneous analysis of these systems demonstrates the ability of SPARC4 to deliver high-precision multi-band photometry and improved orbital and physical parameters.

%% Please use the acknowledgment and contribution environments. This will 
%% be anonomyized when the "anonymous" style option is used. 
\begin{acknowledgments}
This paper uses data obtained with the Simultaneous Polarimeter and Rapid Camera in 4 bands (SPARC4), installed on the 1.6-m telescope at the Observatório do Pico dos Dias (OPD), managed by the Laboratório Nacional de Astrofísica (LNA) under the Ministério da Ciência, Tecnologia e Inovação (Brazil). SPARC4 was funded by Financiadora de Estudos e Projetos - Finep (Proc: 0/1/16/0076/00), Agência Espacial Brasileira - AEB (PO 20VB.0009), Fundação de Amparo à Pesquisa do Estado de São Paulo - FAPESP (Grant 2010/01584-8), Fundação de Amparo à Pesquisa do Estado de Minas Gerais - FAPEMIG (APQ-00193-15 \& APQ-02423-21), Conselho Nacional de Desenvolvimento Científico e Tecnológico - CNPq (Grant \#420812/2018-0) and INCT-Astrofísica. 

E.M. acknowledges funding from Funda\c{c}\~{a}o de Amparo \`{a} Pesquisa do Estado de Minas Gerais (FAPEMIG) under project number APQ-02493-22 and research productivity grant (PQ) number 309829/2022-4 awarded by the National Council for Scientific and Technological Development (CNPq), Brazil.

F.M. thanks the financial support given by the São Paulo Research Foundation (FAPESP), Brazil, Process Number 2024/16260-6.

C.V.R. thanks the Brazilian Space Agency (AEB) by the support from PO 20VB.0009 and the Brazilian National Council for Scientific and Technological Development – CNPq (Proc: 305991/2024-8).

D.V.B. and G.H.S.S. thank to Coordenação de Aperfeiçoamento de Pessoal de Nível Superior - CAPES (Proc: 88887.513623/2020-00, 88887.952075/2024-00) for the scholarship funding.

Funding for the TESS mission is provided by NASA’s Science Mission Directorate. We acknowledge the use of public TESS data from pipelines at the TESS Science Office and at the TESS Science Processing Operations Center. Resources supporting this work were provided by the NASA High-End Computing (HEC) Program through the NASA Advanced Supercomputing (NAS) Division at Ames Research Center for the production of the SPOC data products. DR was supported by NASA under award number NNA16BD14C for NASA Academic Mission Services. TESS data presented in this paper were obtained from the Mikulski Archive for Space Telescopes (MAST) at the Space Telescope Science Institute.

This work has made use of data from the European Space Agency (ESA) mission Gaia (\url{https://www.cosmos.esa.int/gaia}), processed by the Gaia Data Processing and Analysis Consortium (DPAC, \url{https://www.cosmos.esa.int/web/gaia/dpac/consortium}). Funding for the DPAC has been provided by national institutions, in particular, the institutions participating in the Gaia Multilateral Agreement.

\end{acknowledgments}

%\begin{contribution}
%%This section gives authors the space to recognize author contributions. The text inside this environment is NOT counted towards the total word quanta. At a minimum, manuscripts are expected to include this text:

%All editors contribute equally to the operation of PASP.

%% But authors are expected to provide more specific details, e.g. 
%%
%%SC was responsible for writing and submitting the manuscript.
%%WWM came up with the initial research concept and edited the manuscript.
%%OTS obtained the funding and edited the manuscript.
%%EBF provided the formal analysis and validation. He also edited the manuscript.
%%GEH Supervised the undergraduates, wrote the software and administers the project github and Zenodo repositories.
%%
%% Authors can use the Contributor Role Taxonomy (CRediT) at
%% for ideas on how write a good statement tailored to their needs.

%\end{contribution}

\facilities{TESS, Kepler:K2, OPD:PE1.6m:SPARC4}

\software{astropop \citep{Campagnolo2019}, astropy \citep{2013A&A...558A..33A,2018AJ....156..123A,2022ApJ...935..167A}, astroquery \citep{Ginsburg2019}, emcee \citep{foreman2013}, lightkurve \citep{Lightkurve2013}, matplotlib \citep{Hunter2007}, numpy \citep{VanDerWalt2011}, photutils \citep{larry_bradley_2024_13989456}, sparc4-pipeline \citep{Martioli2025}}


\appendix

%-------------------------------------------------------------------
\section{SPARC4 Pipeline data products}
\label{app:s4products}

The results of the SPARC4 Pipeline reduction are saved as FITS data products, which are described in the following sections.

\subsection{Master Calibration products}
\label{sec:mastercalibproducts}

The Master Calibration products (\texttt{*Master\{\$TYPE\}.fits}) are FITS files with an image extension containing calibration data derived from a statistical combination of multiple exposures of the same calibration type. These types are zero, dome flat, sky flat, or dark frames. 

\subsection{Science Image product}
\label{sec:sciimgproducts}

The Science Image product (\texttt{*proc.fits} or \texttt{*stack.fits}) is a FITS file containing a science image that has undergone zero subtraction, flat-fielding, and gain correction. The data may correspond either to a single frame or to a stack of multiple frames. The FITS header documents the reduction process and includes the astrometric calibration, provided through WCS standard keywords \citep{Greisen2002,Calabretta2002}.  In addition to the image extension, the FITS file includes one or more source catalogs as separate extensions. Each catalog corresponds to a specific aperture radius used in the photometry; in the case of polarimetry, separate catalogs are provided for each component of the dual-beam image. Table~\ref{tab:sci_img_catalog_quantities} summarizes the quantities stored in a FITS table extension of the Science Image product.

\begin{deluxetable}{lcll}
\tablecaption{Quantities stored in the catalog FITS table extension of a Science Image product.
\label{tab:sci_img_catalog_quantities}}
\tablehead{
\colhead{Name} & \colhead{Type} & \colhead{Unit} & \colhead{Description}
}
\startdata
SRCINDEX & integer & --    & Source index \\
RA       & float & deg   & Right ascension (J2000) \\
DEC      & float & deg   & Declination (J2000) \\
X        & float & pixel & $x$ coordinate \\
Y        & float & pixel & $y$ coordinate \\
FWHMX   & float & pixel & FWHM along $x$ axis \\
FWHMY   & float & pixel & FWHM along $y$ axis \\
MAG     & float & mag   & Instrumental magnitude \\
EMAG    & float & mag   & Magnitude uncertainty \\
SKYMAG  & float & mag   & Instrumental sky magnitude \\
ESKYMAG & float & mag   & Sky magnitude uncertainty \\
APER    & float & pixel & Aperture radius \\
FLAG    & integer & --    & Photometry control flag \\
\enddata
\end{deluxetable}

\subsection{Polarimetry product}
\label{sec:polarproduct}

The Polarimetry product (\texttt{*polar.fits}) is a FITS file containing several table extensions with polarimetric results for all detected sources. Each extension corresponds to a specific aperture radius used in the photometry.

The structure of each FITS table is similar to that of the catalogs in the Science Image products, but with additional polarimetric information. Specifically, the tables include the Stokes parameters ($Q$, $U$, $V$) and their uncertainties, the total polarization and its angle, the normalization constant, the waveplate zero position, the number of observations, the number of fitted parameters, the chi-square of the fit, and a control polarization flag. In addition, the tables provide the measured values and uncertainties of the flux difference normalized by the sum of the ordinary and extraordinary beam fluxes (Equation \ref{eq:zpol}) for all waveplate position angles obtained during the polarimetric sequence. Table~\ref{tab:polar_catalog_quantities} summarizes the quantities stored in a FITS table extension of the Polarimetry product.

\begin{deluxetable}{lcll}
\tablecaption{Quantities stored in the catalog FITS table extension of the Polarimetry product.
\label{tab:polar_catalog_quantities}}
\tablehead{
\colhead{Name} & \colhead{Type} & \colhead{Unit} & \colhead{Description}
}
\startdata
APERINDEX  & integer & --    & Aperture index \\
APER       & float & pixel & Aperture radius \\
SRCINDEX   & integer & --    & Source index \\
RA         & float & deg   & Right ascension (J2000) \\
DEC        & float & deg   & Declination (J2000) \\
X1         & float & pixel & $x$ coordinate in ordinary beam \\
Y1         & float & pixel & $y$ coordinate in ordinary beam \\
X2         & float & pixel & $x$ coordinate in extraordinary beam \\
Y2         & float & pixel & $y$ coordinate in extraordinary beam \\
FWHM       & float & pixel & Mean FWHM of source \\
MAG        & float & mag   & Instrumental magnitude \\
EMAG       & float & mag   & Magnitude uncertainty \\
SKYMAG     & float & mag   & Instrumental sky magnitude \\
ESKYMAG    & float & mag   & Sky magnitude uncertainty \\
PHOTFLAG   & float & --    & Photometry control flag \\
Q          & float & --    & Stokes parameter $Q$ \\
EQ         & float & --    & Uncertainty in $Q$ \\
U          & float & --    & Stokes parameter $U$ \\
EU         & float & --    & Uncertainty in $U$ \\
V          & float & --    & Stokes parameter $V$ \\
EV         & float & --    & Uncertainty in $V$ \\
P          & float & --    & Total polarization \\
EP         & float & --    & Uncertainty in total polarization \\
THETA      & float & deg   & Polarization angle \\
ETHETA     & float & deg   & Uncertainty in polarization angle \\
K          & float & --    & Normalization constant \\
EK         & float & --    & Uncertainty in normalization \\
ZERO       & float & deg   & Waveplate zero position \\
EZERO      & float & deg   & Uncertainty in waveplate zero \\
NOBS       & integer & --    & Number of observations \\
NPAR       & integer & --    & Number of fitted parameters \\
CHI2       & float & --    & Chi-square of the polarimetric fit \\
RMS        & float & --    & RMS of residuals \\
TSIGMA     & float & --    & Theoretical sigma \\
POLARFLAG  & integer & --    & Polarimetry control flag \\
FOxxxx, EFOxxxx, FExxxx, EFExxxx & float & -- & Flux difference measurements for each waveplate angle (see text) \\
\enddata
\tablecomments{Columns FOxxxx, EFOxxxx, FExxxx, and EFExxxx are provided for all waveplate position angles. The `xxxx` suffix represents the waveplate step index.}
\end{deluxetable}

\subsection{Light Curve product}
\label{sec:lcproduct}

The Light Curve product (\texttt{*lc.fits}) is a FITS file containing photometric catalogs compiled from a series of Science Image products. It consists of several table extensions, each corresponding to a specific aperture radius used in the photometry.

Each FITS table extension includes the same data as the Science Image catalogs, with the addition of the mid-exposure time (BJD), the RMS of the median magnitude dispersion computed within 10-minute windows, and extra keywords specified by the user in the pipeline parameter file. For example, one may add the airmass (\texttt{AIRMASS}) or any other value stored as a FITS header keyword. Table~\ref{tab:lc_catalog_quantities} summarizes the quantities stored in a FITS table extension of the Light Curve product.

\begin{deluxetable}{lcll}
\tablecaption{Quantities stored in the FITS table extension of a Light Curve product.
\label{tab:lc_catalog_quantities}}
\tablehead{
\colhead{Name} & \colhead{Type} & \colhead{Unit} & \colhead{Description}
}
\startdata
TIME      & float & BJD      & Mid-exposure time \\
SRCINDEX  & integer & --       & Source index \\
RA        & float & deg      & Right ascension (J2000) \\
DEC       & float & deg      & Declination (J2000) \\
X         & float & pixel    & $x$ coordinate \\
Y         & float & pixel    & $y$ coordinate \\
FWHM      & float & pixel    & Mean FWHM of source \\
MAG       & float & mag      & Instrumental magnitude \\
EMAG      & float & mag      & Magnitude uncertainty \\
SKYMAG    & float & mag      & Instrumental sky magnitude \\
ESKYMAG   & float & mag      & Sky magnitude uncertainty \\
FLAG      & integer   & --       & Photometry control flag \\
RMS       & float & mag      & RMS of median magnitude within 10-min windows \\
AIRMASS   & float & --       & Airmass (optional, user-defined) \\
\enddata
\end{deluxetable}

\subsection{Time Series product}
\label{sec:tsproduct}

The Time Series product (\texttt{*ts.fits}) is a FITS file compiling the information contained in the catalog extensions of a series of Polarimetry products. It provides time-resolved polarimetric measurements for all sources observed.

Each FITS table includes astrometric, photometric, and polarimetric data, together with their associated uncertainties. In particular, the tables contain the Stokes parameters ($Q$, $U$, $V$), the total polarization and polarization angle, the normalization constant, the waveplate zero position, the number of observations, the number of fitted parameters, and the chi-square of the fit. Table~\ref{tab:ts_catalog_quantities} summarizes the quantities stored in a FITS table extension of the Time Series product.

\begin{deluxetable}{lcll}
\tablecaption{Quantities stored in the FITS table extension of a Time Series product.
\label{tab:ts_catalog_quantities}}
\tablehead{
\colhead{Name} & \colhead{Type} & \colhead{Unit} & \colhead{Description}
}
\startdata
TIME      & float & BJD      & Mid-exposure time \\
SRCINDEX  & integer   & --       & Source index \\
RA        & float & deg      & Right ascension (J2000) \\
DEC       & float & deg      & Declination (J2000) \\
X1        & float & pixel    & $x$ coordinate in ordinary beam \\
Y1        & float & pixel    & $y$ coordinate in ordinary beam \\
X2        & float & pixel    & $x$ coordinate in extraordinary beam \\
Y2        & float & pixel    & $y$ coordinate in extraordinary beam \\
FWHM      & float & pixel    & Mean FWHM of source \\
MAG       & float & mag      & Instrumental magnitude \\
EMAG      & float & mag      & Magnitude uncertainty \\
Q         & float & --       & Stokes parameter $Q$ \\
EQ        & float & --       & Uncertainty in $Q$ \\
U         & float & --       & Stokes parameter $U$ \\
EU        & float & --       & Uncertainty in $U$ \\
V         & float & --       & Stokes parameter $V$ \\
EV        & float & --       & Uncertainty in $V$ \\
P         & float & --       & Total polarization \\
EP        & float & --       & Uncertainty in total polarization \\
THETA     & float & deg      & Polarization angle \\
ETHETA    & float & deg      & Uncertainty in polarization angle \\
K         & float & --       & Normalization constant \\
EK        & float & --       & Uncertainty in normalization constant \\
ZERO      & float & deg      & Waveplate zero position \\
EZERO     & float & deg      & Uncertainty in waveplate zero position\\
NOBS      & integer   & --       & Number of observations \\
NPAR      & integer   & --       & Number of fitted parameters \\
CHI2      & float & --       & Chi-square of the polarimetric fit \\
\enddata
\end{deluxetable}

%-------------------------------------------------------------------
\section{Photometric calibration of SPARC4 Pipeline data}
\label{app:photometricCalibration}

An initial photometric characterization of the SPARC4 instrument was presented by \citet{Schlindwein2024}, who adopted a simple model to convert SPARC4 magnitudes to a standard photometric system. Their approach is useful for calibrating data from a single pointing field but does not account for the dependence on the airmass of the observations. We propose an approach here that follows the methodology described in \cite{Jablonski1994}, based on the fundamental concepts introduced by \cite{Harris1981}. In this analysis, we extend the literature methods to multi-band and multi-object simultaneous observations spanning hours-long time series over several nights, reflecting the characteristics of the SPARC4 transit data analyzed here.  

\begin{figure*}
   \centering
       \includegraphics[width=1.\hsize]{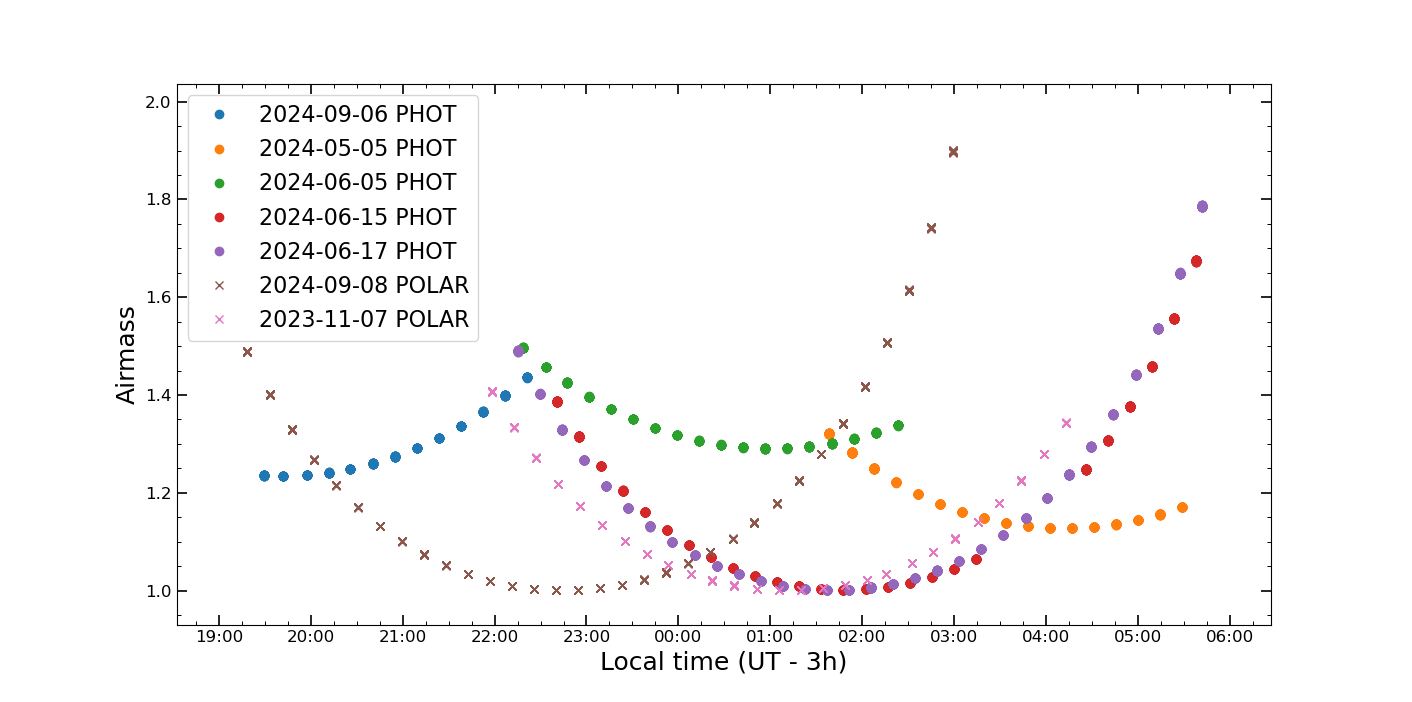}
      \caption{Air mass of observations as a function of OPD local time (UT - 3h). Crosses represent observations in POLAR mode, and circles represent observations in PHOT mode. The dates of the observations are indicated in the legend.
      }
      \label{fig:airmass_coverage}
  \end{figure*}

We use the \emph{g}, \emph{r}, \emph{i}, and \emph{z} magnitudes from SkyMapper, converted to the SDSS system \citep{Wolf2018}, as references to calibrate our data. The WASP-123 field has no observations in SkyMapper and is therefore excluded from this analysis. The basic idea for obtaining an absolute photometric calibration is to solve a least-squares (LS) minimization problem constrained by the following equations:

\begin{align}
\label{eq:photcalibeqs1}
g'-g &=g_{0} + g_{1}(g-r) + g_{2}X, \\ 
\label{eq:photcalibeqs2}
r'-r &=r_{0} + r_{1}(r-i) + r_{2}X, \\ 
\label{eq:photcalibeqs3}
i'-i &=i_{0} + i_{1}(i-z) + i_{2}X, \\ 
\label{eq:photcalibeqs4}
z'-z &=z_{0} + z_{1}(i-z) + z_{2}X, \\ 
\label{eq:photcalibeqs5}
g'-r' &=a_{0} + a_{1}(g-r) + a_{2}X, \\ 
\label{eq:photcalibeqs6}
r'-i' &=b_{0} + b_{1}(r-i) + b_{2}X, \\ 
\label{eq:photcalibeqs7}
i'-z' &=c_{0} + c_{1}(i-z) + c_{2}X.  
\end{align}

\noindent where $g,r,i,z$ represent the reference magnitudes, and the corresponding primed values $g',r',i',z'$ are the instrumental magnitudes. The coefficients $g_{j}, r_{j}, i_{j}, z_{j}, a_{j}, b_{j}, c_{j}$ for $j=\{0,1,2\}$ are system transformation (0 and 1) and extinction coefficients (2); and $X$ is the airmass of observations. Figure~\ref{fig:airmass_coverage} shows the airmasses of our observations, which range from 1.0 to 1.9. The extinction coefficients presented here are valid within these limits.

Given that our observations consist of high-cadence time series spanning several hours, we first bin the instrumental magnitudes in 15-minute intervals before solving Eqs. \ref{eq:photcalibeqs1} -- \ref{eq:photcalibeqs7}. This step mitigates high-frequency noise caused by atmospheric scintillation and local seeing variations. We also discarded HATS-24 data with ${\rm BJD} > 2460467.7276$, as conditions were no longer photometric. We also excluded the range $2460477.7611 < {\rm BJD} < 2460477.8092$ (HATS-9, 2024-06-15) due to a dome tracking failure that partially obstructed the telescope’s field of view.

\begin{figure*}
\centering
\includegraphics[width=1.0\hsize]{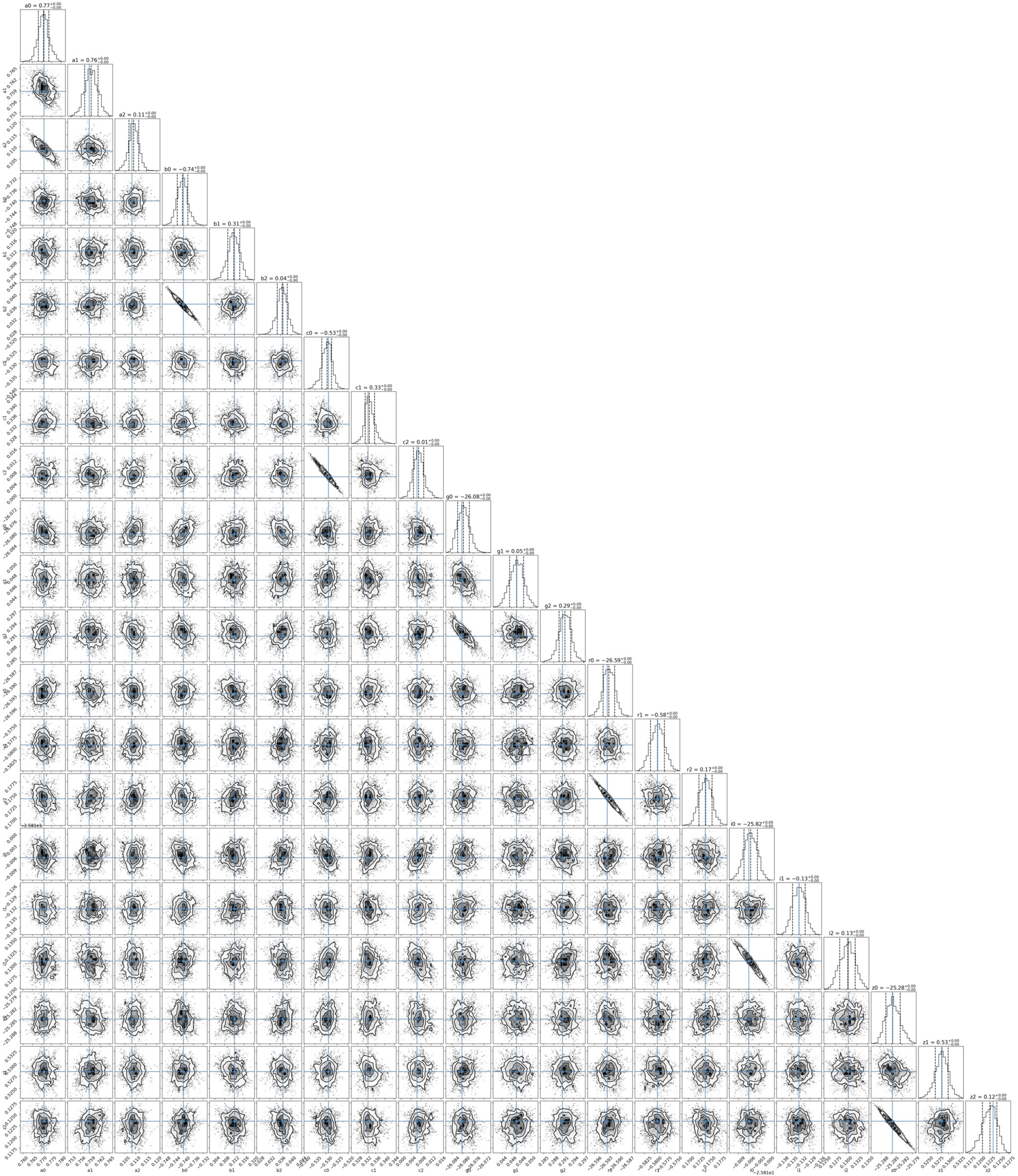}
\caption{Pairs plot showing the MCMC samples and posterior distributions of the coefficients obtained in PHOT mode on the night of 20240617, from the time series of the HATS-9 field.  The contours mark the 1$\sigma$, 2$\sigma$, and 3$\sigma$ regions of the distribution. The blue crosses indicate the best-fit values for each parameter obtained by the mode, and the dashed vertical lines in the projected distributions show the median values and the 1$\sigma$ uncertainty (34\% on each side of the median).}
\label{fig:hats9photcoeffs_pairsplot}
\end{figure*}

\begin{figure*}
\centering
\includegraphics[width=1.0\hsize]{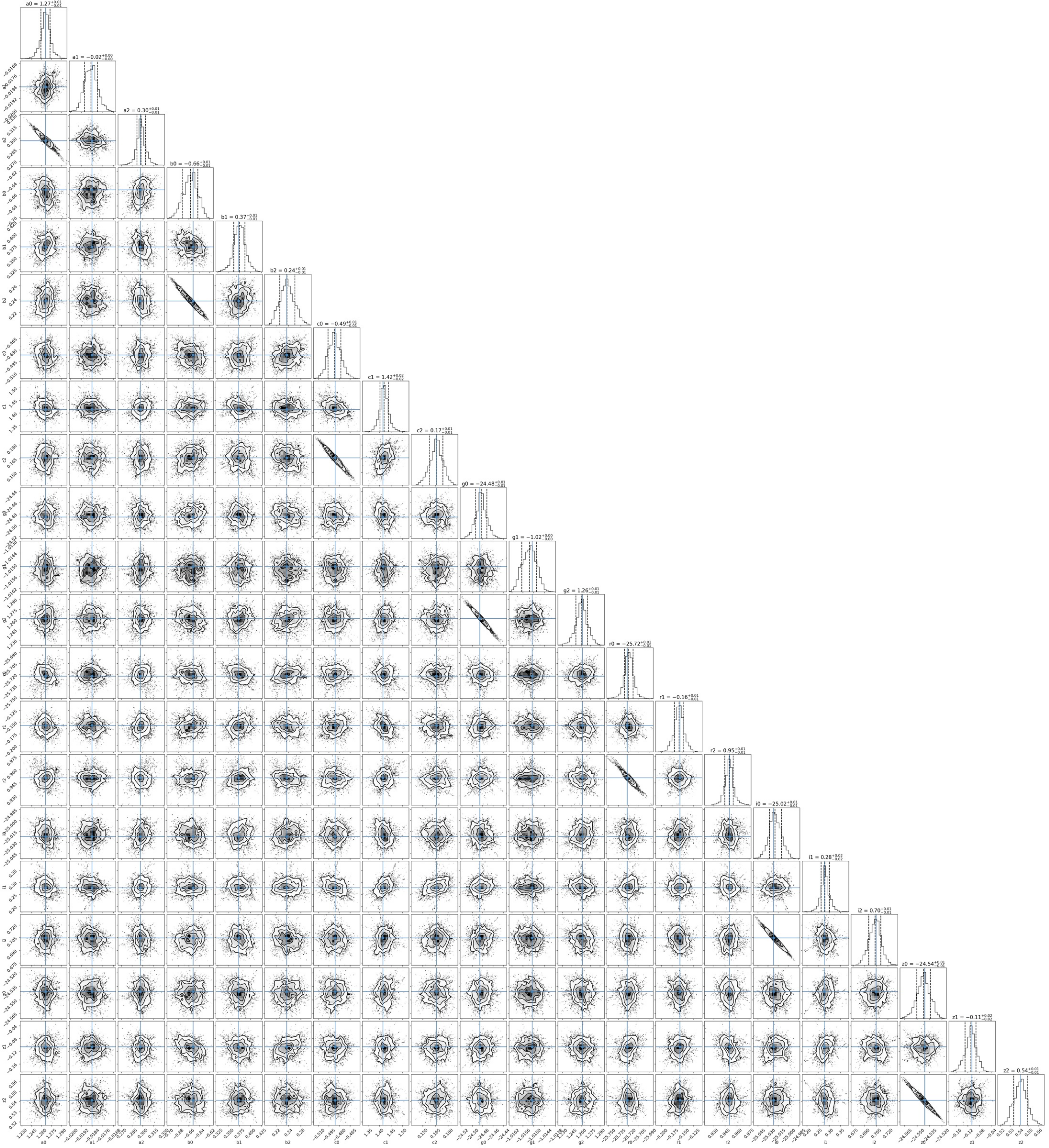}
\caption{Same as Figure~\ref{fig:hats9photcoeffs_pairsplot} but for data obtained in POLAR mode on the night of 20240908, from the time series of the WASP-111 field.}
\label{fig:wasp111photcoeffs_pairsplot}
\end{figure*}

We used the \texttt{scipy.optimize.leastsq} tool to fit the coefficients independently for each night. In addition, we run a Bayesian Markov Chain Monte Carlo (MCMC) analysis using \texttt{emcee} \citep{foreman2013}, where we adopted uniform priors for the coefficients, with initial values set by the LS solution.  Each MCMC run adopts 52 walkers, 100 burn-in iterations followed by 500 iterations. Figures \ref{fig:hats9photcoeffs_pairsplot} and \ref{fig:wasp111photcoeffs_pairsplot} show two examples of corner plots of the posterior distributions of the coefficients for HATS-9/2024-06-17/PHOT mode and WASP-111/2024-09-08/POLAR mode. The results are summarized in Table \ref{tab:photcalibresults}.   There is no significant offset between the coefficients obtained in PHOT and POLAR modes, indicating that the polarimetric optics do not affect strongly the flux calibration. Though, it would be valuable to repeat this analysis using observations that alternate between the two modes while pointing at the same object under nearly identical time and airmass conditions.  

We used the nightly coefficients from Table~\ref{tab:photcalibresults} to calibrate the instrumental magnitudes and colors of all sources of the same night. We then estimated the photometric precision by computing the median absolute deviation (MAD) of the differences between the calibrated values and the reference catalog values. As an example, in the night of 2024-06-17 in the field of HATS-9 observed in PHOT mode, we obtained a precision of $\sigma_{g-r}=0.28$~mag, $\sigma_{r-i}=0.21$~mag, and $\sigma_{i-z}=0.28$~mag for the colors and $\sigma_{g}=0.12$~mag, $\sigma_{r}=0.11$~mag, $\sigma_{i}=0.12$~mag, and $\sigma_{z}=0.15$~mag for the magnitudes. For the other nights, we obtained similar dispersions, both in PHOT and POLAR modes. 

\begin{table*}[ht!]
\centering
%\tiny
\scriptsize
\caption{Measured values of the coefficients from the absolute photometric calibration of SPARC4, based on observations of seven transits of exoplanets — five obtained in PHOT mode and two in POLAR mode.}
\label{tab:photcalibresults} 
\begin{tabular}{cccccccc}
\hline\hline
Night & $g_{0}$ & $r_{0}$ & $i_{0}$ & $z_{0}$ & $a_{0}$ & $b_{0}$ & $c_{0}$\\
\hline
20231107 &  $-23.315\pm0.002$ & $-24.449\pm0.003$ & $-23.320\pm0.003$ & $-22.755\pm0.003$ & $1.137\pm0.004$ & $-1.124\pm0.004$ & $-0.556\pm0.004$  \\
20240505 &  $-24.445\pm0.004$ & $-25.131\pm0.004$ & $-24.209\pm0.005$ & $-23.757\pm0.004$ & $0.708\pm0.005$ & $-0.865\pm0.006$ & $-0.429\pm0.005$  \\
20240605 &  $-24.511\pm0.010$ & $-24.601\pm0.013$ & $-23.872\pm0.009$ & $-23.601\pm0.011$ & $0.912\pm0.008$ & $-0.587\pm0.008$ & $-0.392\pm0.010$  \\
20240615 &  $-26.003\pm0.001$ & $-26.669\pm0.001$ & $-25.765\pm0.001$ & $-25.224\pm0.001$ & $0.723\pm0.002$ & $-0.849\pm0.001$ & $-0.537\pm0.002$  \\
20240617 &  $-26.080\pm0.002$ & $-26.593\pm0.002$ & $-25.815\pm0.002$ & $-25.284\pm0.002$ & $0.771\pm0.002$ & $-0.741\pm0.003$ & $-0.530\pm0.002$  \\
20240906 &  $-24.275\pm0.040$ & $-25.350\pm0.047$ & $-24.787\pm0.025$ & $-24.374\pm0.027$ & $1.083\pm0.052$ & $-0.526\pm0.030$ & $-0.372\pm0.035$  \\
20240908 &  $-24.486\pm0.011$ & $-25.722\pm0.006$ & $-25.022\pm0.009$ & $-24.541\pm0.008$ & $1.266\pm0.007$ & $-0.652\pm0.014$ & $-0.486\pm0.009$  \\
\hline
Median &  $\bf-24.486\pm0.312$ & $\bf-25.350\pm1.109$ & $\bf-24.787\pm1.357$ & $\bf-24.374\pm1.147$ & $\bf0.912\pm0.280$ & $\bf-0.741\pm0.184$ & $\bf-0.486\pm0.084$  \\
\hline\hline
Night & $g_{1}$ & $r_{1}$ & $i_{1}$ & $z_{1}$ & $a_{1}$ & $b_{1}$ & $c_{1}$\\
\hline
20231107 &  $0.019\pm0.001$ & $-0.015\pm0.002$ & $-0.055\pm0.003$ & $-0.129\pm0.003$ & $1.021\pm0.001$ & $1.023\pm0.003$ & $1.065\pm0.004$  \\
20240505 &  $0.133\pm0.002$ & $0.041\pm0.003$ & $-0.101\pm0.006$ & $0.329\pm0.004$ & $1.032\pm0.003$ & $0.881\pm0.005$ & $0.569\pm0.006$  \\
20240605 &  $-0.009\pm0.002$ & $-0.635\pm0.006$ & $-0.431\pm0.004$ & $0.656\pm0.004$ & $0.520\pm0.003$ & $0.329\pm0.006$ & $-0.080\pm0.006$  \\
20240615 &  $0.105\pm0.001$ & $-0.169\pm0.001$ & $-0.125\pm0.002$ & $-0.080\pm0.002$ & $1.076\pm0.001$ & $0.655\pm0.002$ & $0.987\pm0.002$  \\
20240617 &  $0.047\pm0.001$ & $-0.580\pm0.002$ & $-0.133\pm0.002$ & $0.529\pm0.002$ & $0.760\pm0.002$ & $0.311\pm0.002$ & $0.334\pm0.003$  \\
20240906 &  $-0.041\pm0.014$ & $-0.038\pm0.022$ & $-0.002\pm0.003$ & $-0.140\pm0.018$ & $0.967\pm0.018$ & $0.851\pm0.025$ & $1.125\pm0.026$  \\
20240908 &  $-1.015\pm0.000$ & $-0.157\pm0.011$ & $0.278\pm0.019$ & $-0.106\pm0.019$ & $-0.018\pm0.000$ & $0.373\pm0.013$ & $1.414\pm0.018$  \\
\hline
Median &  $\bf0.019\pm0.089$ & $\bf-0.157\pm0.211$ & $\bf-0.101\pm0.068$ & $\bf-0.080\pm0.088$ & $\bf0.967\pm0.161$ & $\bf0.655\pm0.419$ & $\bf0.987\pm0.619$  \\
\hline\hline
Night & $g_{2}$ & $r_{2}$ & $i_{2}$ & $z_{2}$ & $a_{2}$ & $b_{2}$ & $c_{2}$\\
\hline
20231107 &  $0.285\pm0.002$ & $0.189\pm0.003$ & $0.150\pm0.003$ & $0.165\pm0.002$ & $0.096\pm0.003$ & $0.029\pm0.003$ & $-0.022\pm0.004$  \\
20240505 &  $0.129\pm0.004$ & $0.028\pm0.003$ & $0.022\pm0.005$ & $0.089\pm0.003$ & $0.121\pm0.005$ & $-0.007\pm0.005$ & $-0.083\pm0.004$  \\
20240605 &  $0.368\pm0.008$ & $0.085\pm0.010$ & $0.020\pm0.007$ & $0.099\pm0.008$ & $0.117\pm0.006$ & $0.017\pm0.006$ & $0.006\pm0.007$  \\
20240615 &  $0.191\pm0.001$ & $0.116\pm0.001$ & $0.073\pm0.001$ & $0.066\pm0.001$ & $0.078\pm0.002$ & $0.038\pm0.001$ & $0.005\pm0.001$  \\
20240617 &  $0.291\pm0.002$ & $0.174\pm0.001$ & $0.130\pm0.002$ & $0.122\pm0.001$ & $0.109\pm0.002$ & $0.036\pm0.002$ & $0.007\pm0.002$  \\
20240906 &  $0.044\pm0.031$ & $-0.106\pm0.038$ & $-0.046\pm0.020$ & $-0.057\pm0.021$ & $0.130\pm0.042$ & $-0.065\pm0.025$ & $-0.022\pm0.027$  \\
20240908 &  $1.262\pm0.009$ & $0.951\pm0.005$ & $0.705\pm0.008$ & $0.541\pm0.007$ & $0.297\pm0.006$ & $0.237\pm0.012$ & $0.166\pm0.008$  \\
\hline
Median &  $\bf0.285\pm0.139$ & $\bf0.116\pm0.108$ & $\bf0.073\pm0.084$ & $\bf0.099\pm0.049$ & $\bf0.117\pm0.019$ & $\bf0.029\pm0.017$ & $\bf0.005\pm0.039$  \\
\hline

\end{tabular}
\end{table*} 

\begin{figure*}
\centering
    \includegraphics[width=\textwidth]{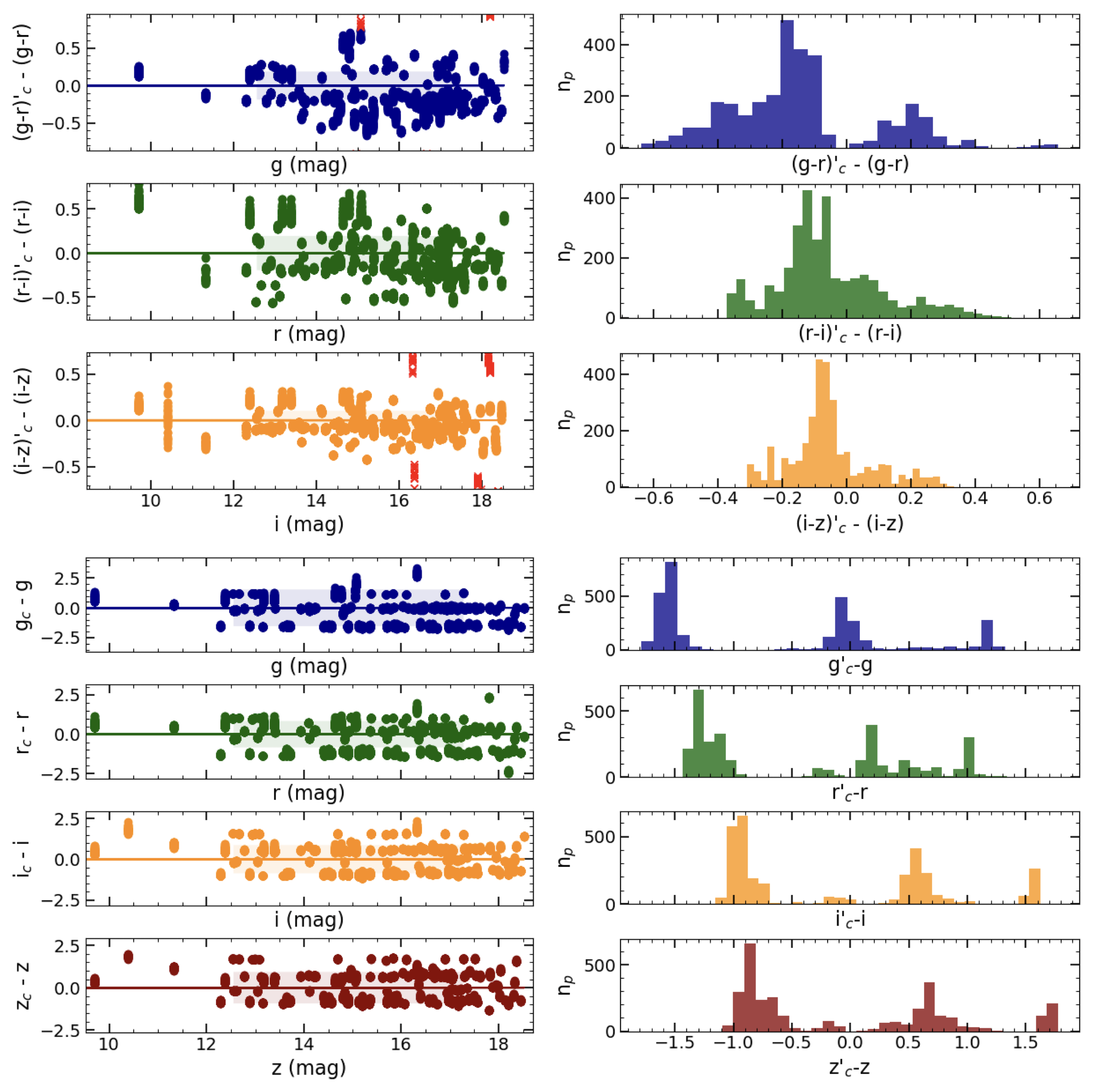}
\caption{Photometric accuracy of SPARC4 absolute photometry calibrated using the master calibration (see text and Table \ref{tab:photcalibresults}). The left panels show the residuals between the calibrated SPARC4 photometry (indicated with the index ``c'') and the SkyMapper reference catalog. Right panels show the histograms of these residuals.  The MAD dispersion of each color index is $\sigma_{g-r}=0.28$~mag, $\sigma_{r-i}=0.27$~mag, and $\sigma_{i-z}=0.16$~mag, based on 6791, 7205, and 6986 measurements, respectively. A sigma-clipping of 5$\sigma$ was applied and the red crosses show the excluded points.} The dispersion of each magnitude shown in the plots above is $\sigma_{g}=1.20$~mag, $\sigma_{r}=0.99$~mag, and $\sigma_{i}=0.94$~mag, $\sigma_{z}=0.90$~mag, based on 6981, 7212, 7232, and 7166 measurements, respectively, after applying a 5$\sigma$ clipping.
\label{fig:photcalib_master}
\end{figure*}

In addition to the nightly calibrations, we computed a master photometric calibration (and its uncertainties) by taking the median and MAD of the nightly coefficients (Table \ref{tab:photcalibresults}). Figure~\ref{fig:photcalib_master} shows the distribution of residuals in the calibrated magnitudes and color indices, derived from all datasets using the master calibration. For this global solution, the accuracy of the magnitudes is limited to approximately $\sim1$~mag, which is significantly worse than the $<0.1$ mag precision typically achieved in same-night photometric calibrations. The histograms clearly exhibit a multimodal distribution, as expected, because the zero-point coefficients ($g_{0}, r_{0}, i_{0}, z_{0}$) are strongly affected by variations in sky transparency and background brightness, which change substantially from night to night at OPD.

In contrast, the color indices are far less sensitive to night-to-night variability, yielding a more precise determination at the level of $\sim0.3$ mag — similar to the precision achieved in same-night calibrations. Therefore, the median coefficients can be used to calibrate observations obtained on regular nights when the available data are insufficient to constrain a robust nightly calibration model, particularly for color indices. Nevertheless, the most accurate absolute photometric calibration is obtained when a calibrator field is observed on the same night under conditions comparable to those of the science target.

%-------------------------------------------------------------------
\section{Polarimetric standards}
\label{app:polarstandards}

\begin{figure*}
\centering
\includegraphics[width=1.\hsize]{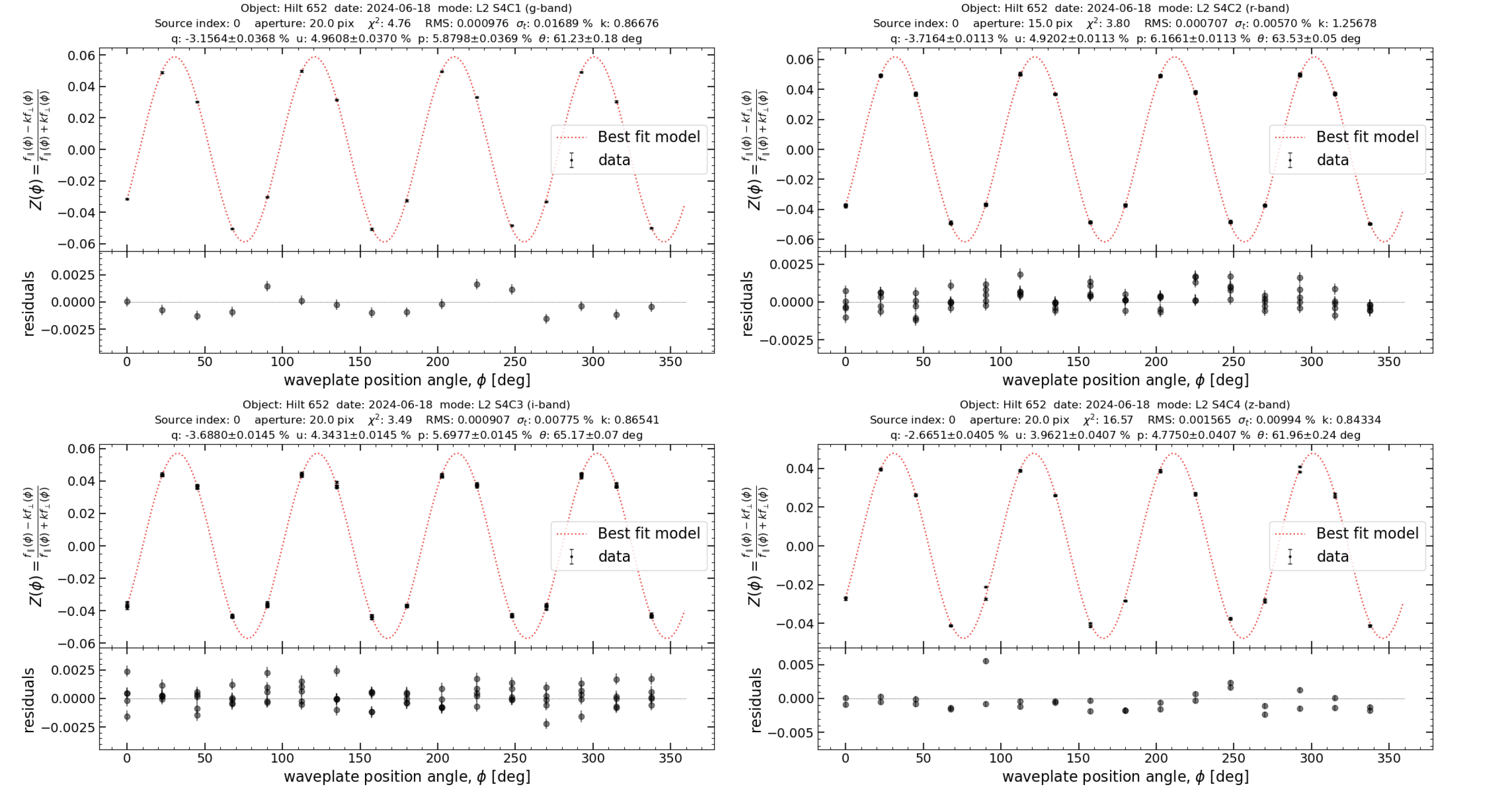}
\caption{Linear polarization of the polarized standard star Hilt~652 measured in the four SPARC4 channels. Model-fit parameters and measurement details are given in each panel header.}
\label{fig:Hilt652_POLAR_L2}
\end{figure*}

\begin{figure*}
\centering
\includegraphics[width=1.\hsize]{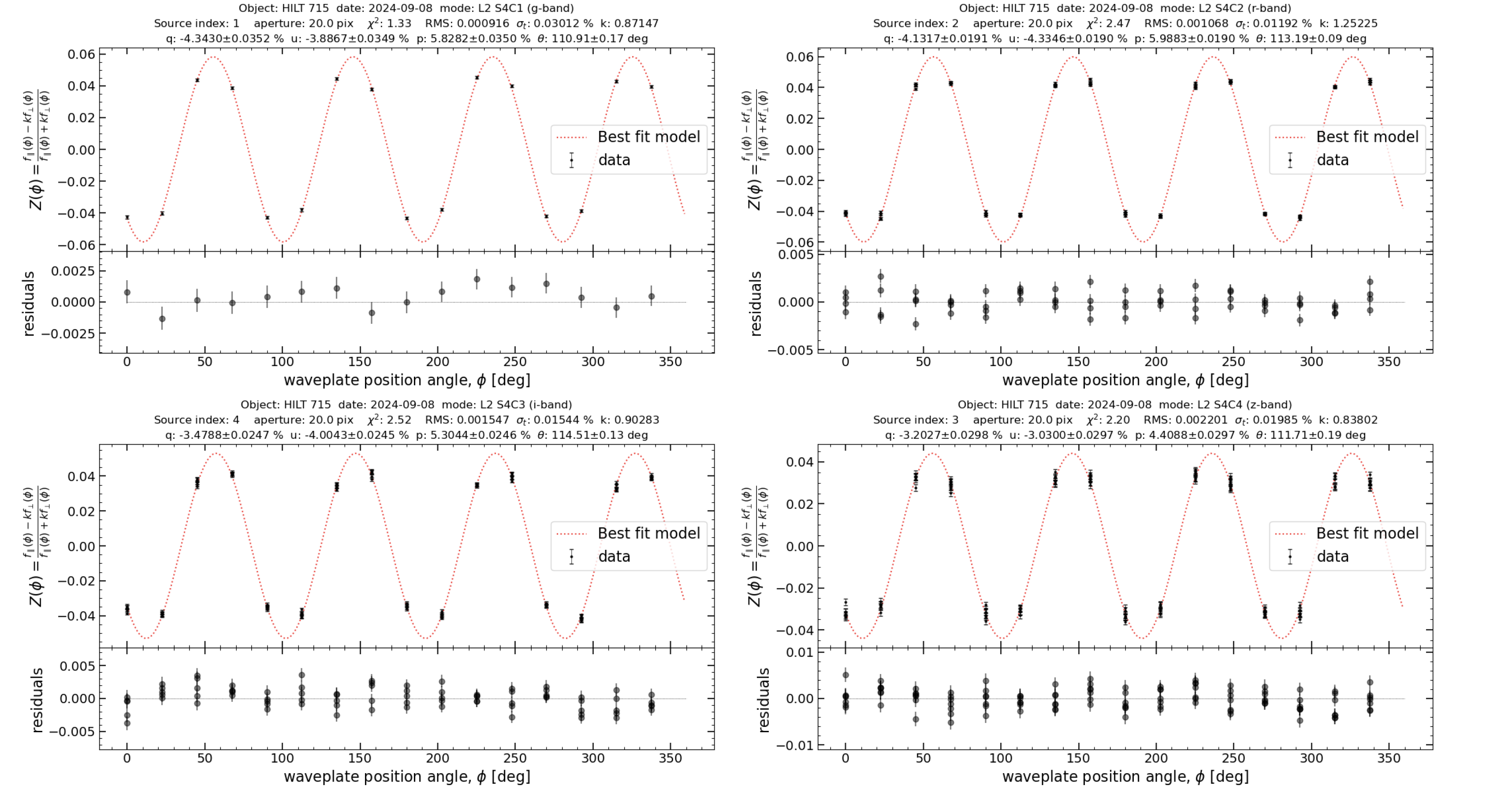}
\caption{Linear polarization of the polarized standard star Hilt~715 measured in the four SPARC4 channels. Model-fit parameters and measurement details are given in each panel header.}
\label{fig:Hilt715_POLAR_L2}
\end{figure*}

\begin{figure*}
\centering
\includegraphics[width=1.\hsize]{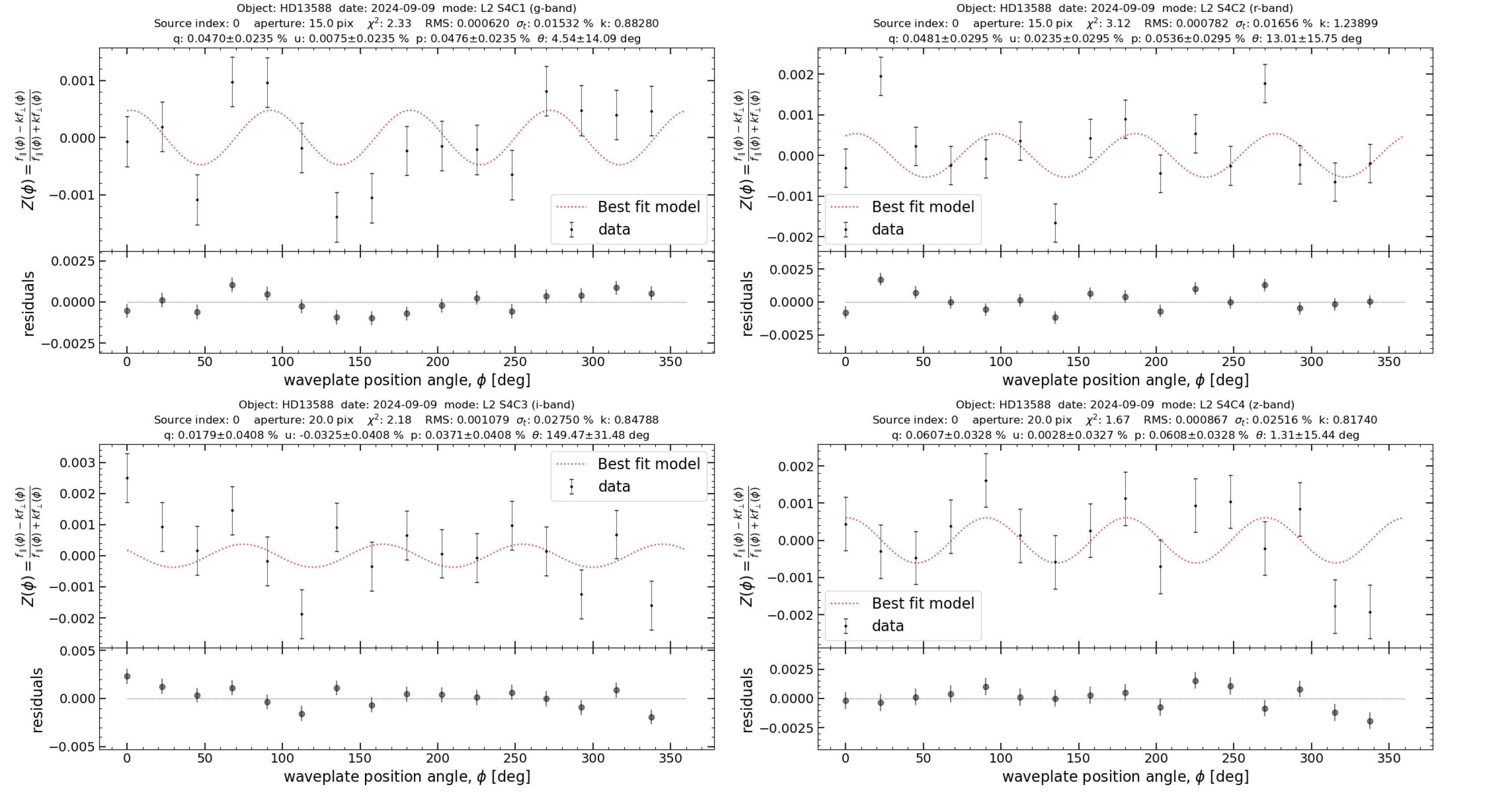}
\caption{Linear polarization of the unpolarized standard star HD~13588 measured in the four SPARC4 channels. Model-fit parameters and measurement details are given in each panel header.}
\label{fig:HD13588_POLAR_L2}
\end{figure*}

In this appendix, we present the pipeline fit that provides the linear polarization of two polarized and one unpolarized standard stars (Figures~\ref{fig:Hilt652_POLAR_L2}, \ref{fig:Hilt715_POLAR_L2}, and \ref{fig:HD13588_POLAR_L2}).

%-------------------------------------------------------------------
\section{TESS, K2, andnSPARC4 light curves and transit fit}
\label{app:s4transitfit}

\begin{figure*}
\centering
\includegraphics[width=0.85\textwidth]{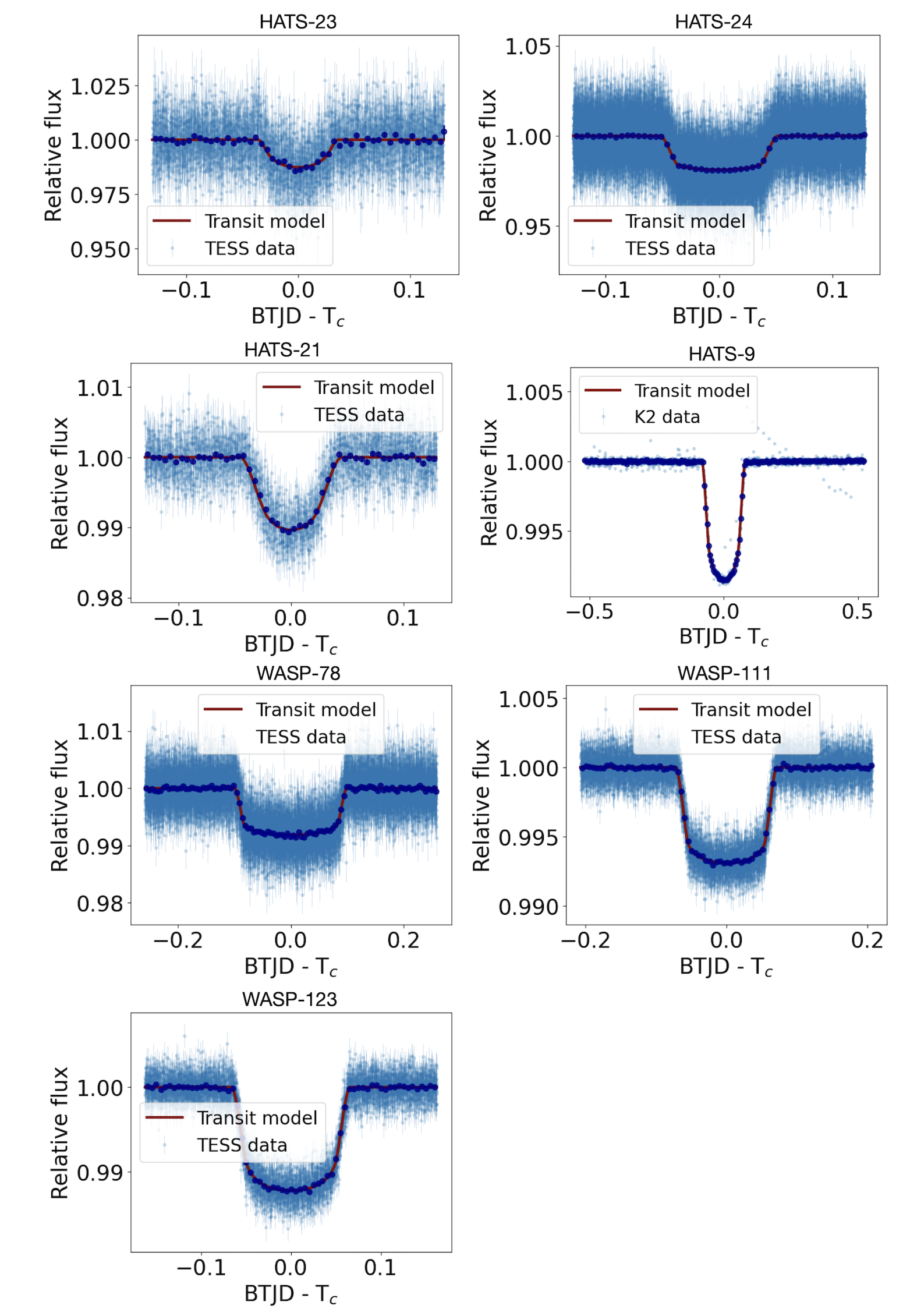}
\caption{TESS (or K2) light curves of the seven exoplanet transits analyzed in this work. Light blue points show the photometric data around the transits, dark blue points show binned data (bin size 0.002~d), and red lines indicate the best-fit joint SPARC4 + TESS (or K2) transit models. Times are relative to the transit midpoint.}
\label{fig:tess_k2_transits}
\end{figure*}

This Appendix present the TESS, K2, and SPARC4 light curves and best-fit transit fit, as detailed in Section \ref{sec:transit-analysis}.  Figure~\ref{fig:tess_k2_transits} presents the phase-folded detrended TESS and K2 light curve data and best-fit model obtained for the seven exoplanets observed with SPARC4.

\begin{figure}
\centering
\includegraphics[width=1.\hsize]{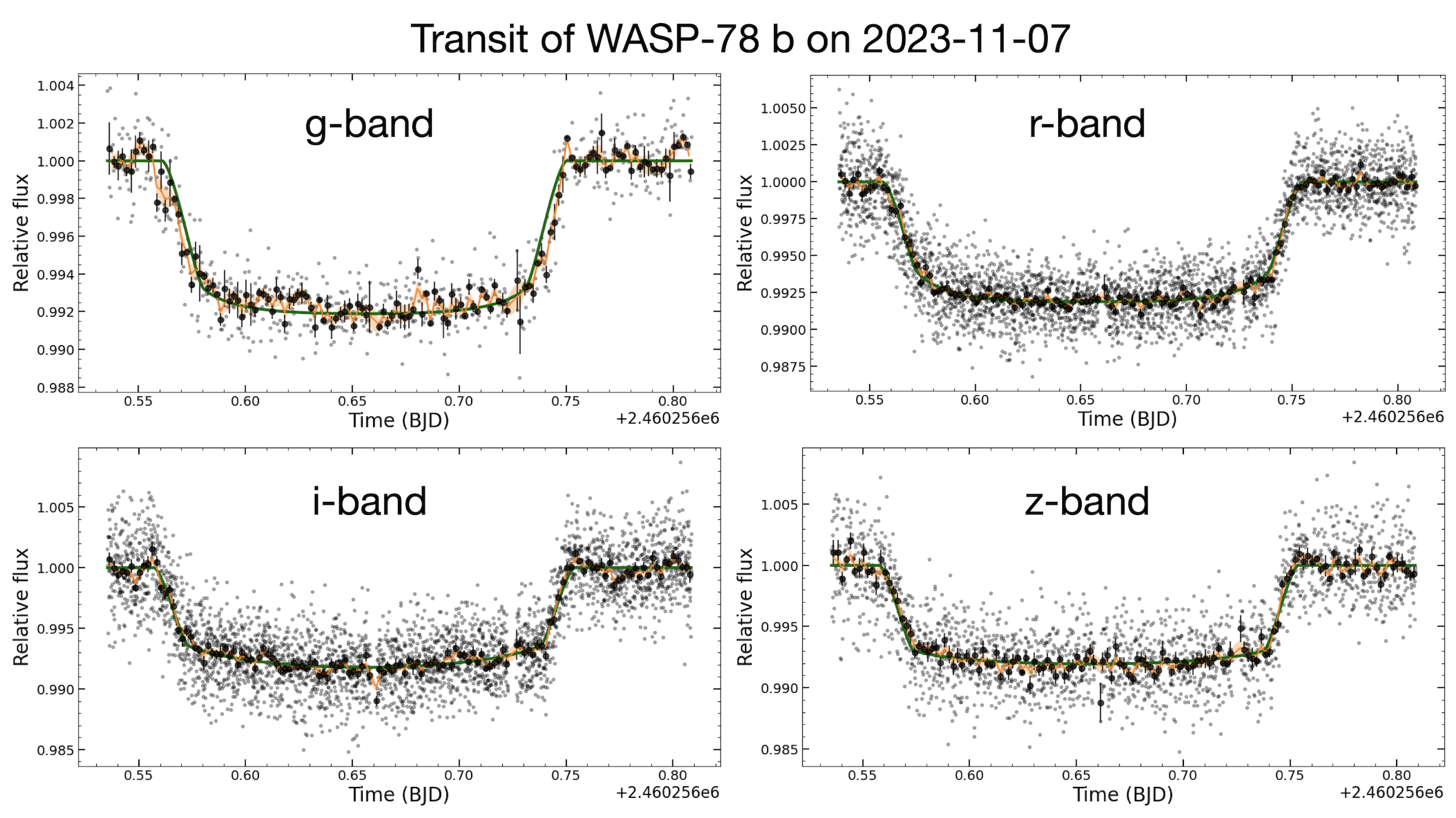}
\caption{Four-band SPARC4 differential photometry time series of the transit of the exoplanet WASP-78~b. Grey points show the observed light curve data, solid black points show binned data (bin size 0.002~d), orange shaded regions show a Gaussian Process interpolation to the binned data, and the green line show the best-fit transit model.}
\label{fig:wasp78_s4_transit}
\end{figure}

\begin{figure}
\centering
\includegraphics[width=1.\hsize]{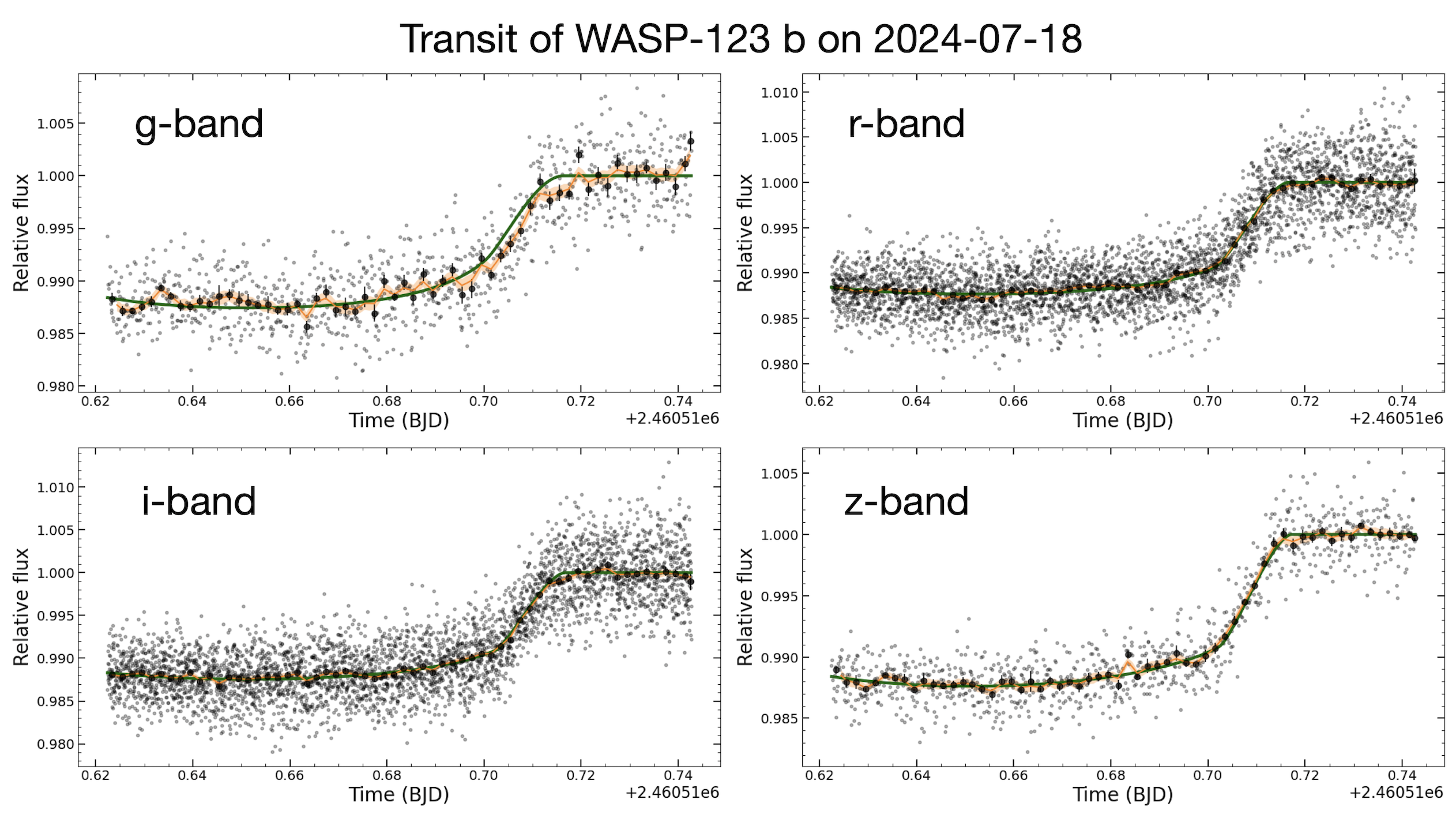}
\caption{Same as Figure~\ref{fig:wasp78_s4_transit}, but for the transit of WASP-123~b.}
\label{fig:wasp123_s4_transit}
\end{figure}

\begin{figure}
\centering
\includegraphics[width=1.\hsize]{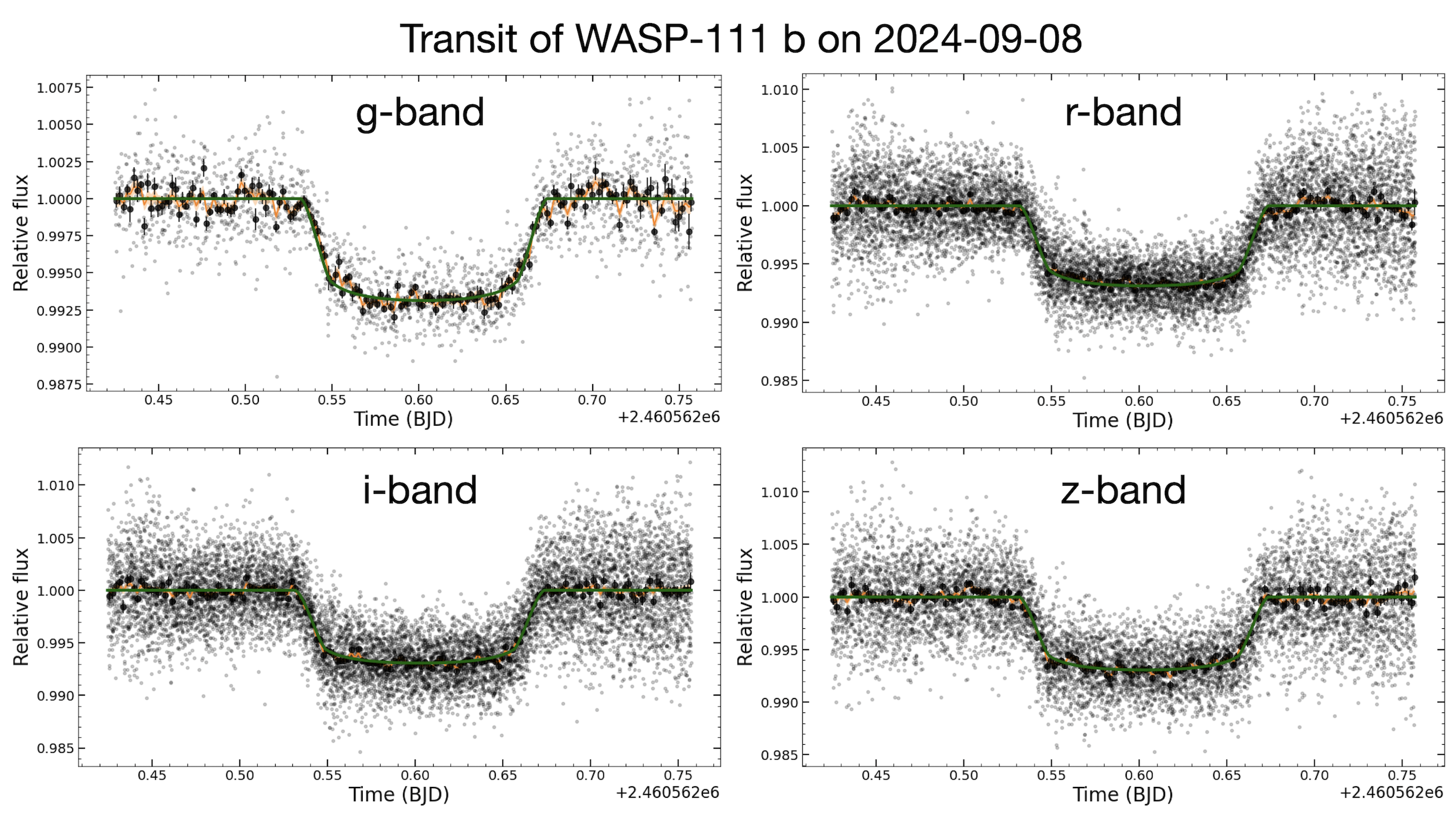}
\caption{Same as Figure~\ref{fig:wasp78_s4_transit}, but for the transit of WASP-111~b.}
\label{fig:wasp111_s4_transit}
\end{figure}

\begin{figure}
\centering
\includegraphics[width=1.\hsize]{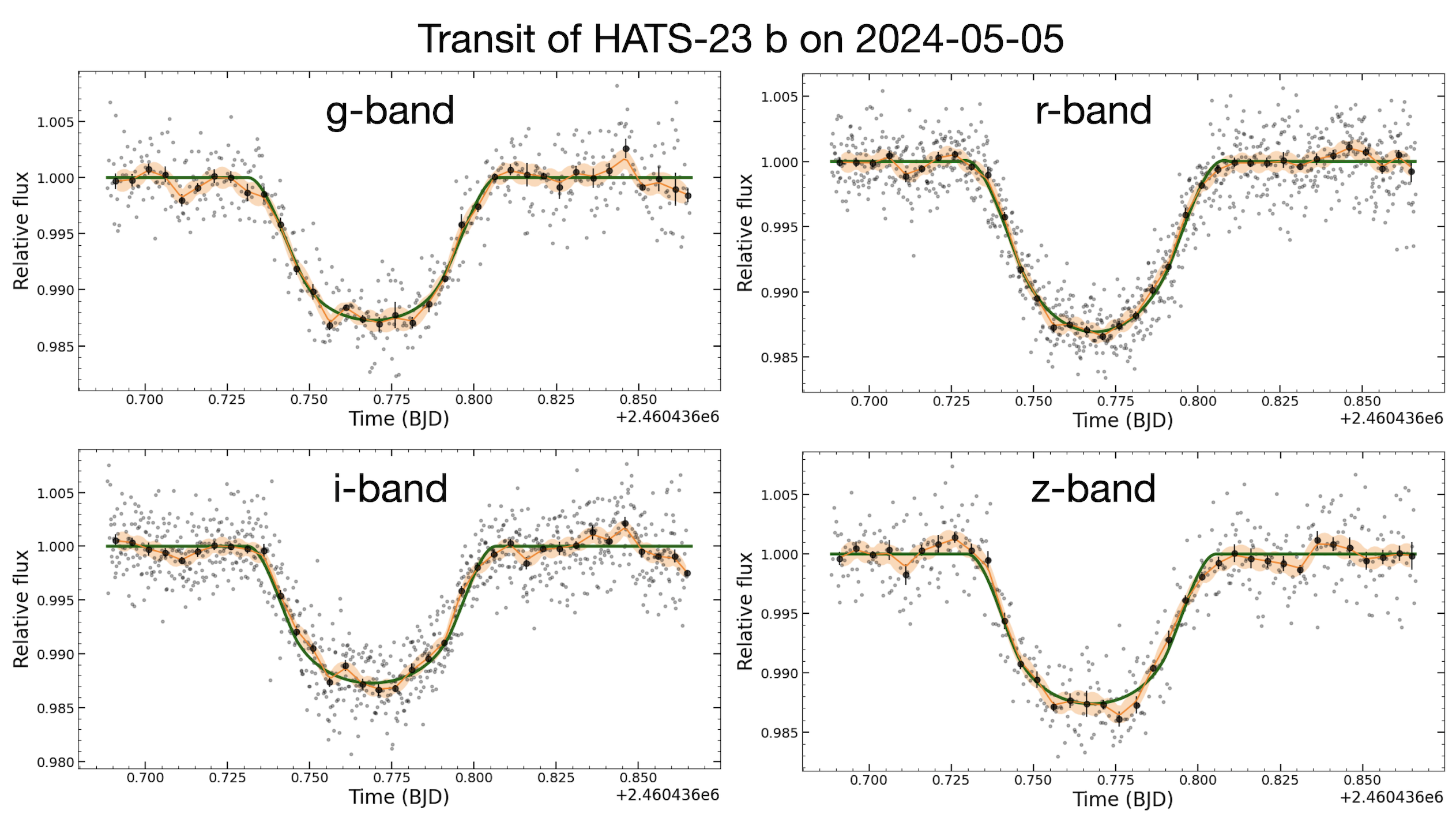}
\caption{Same as Figure~\ref{fig:wasp78_s4_transit}, but for the transit of HATS-23~b.}
\label{fig:hats23_s4_transit}
\end{figure}

\begin{figure}
\centering
\includegraphics[width=1.\hsize]{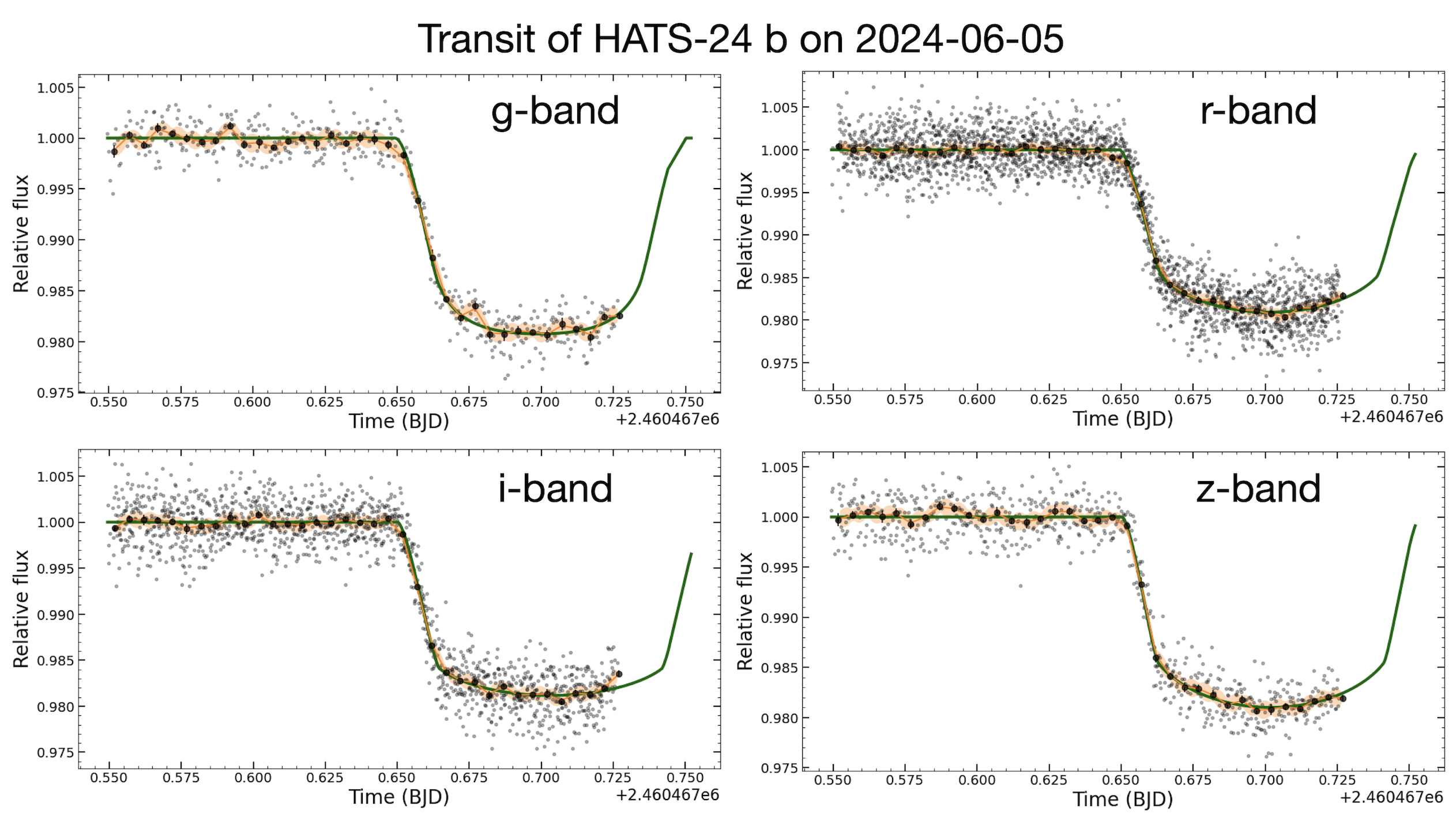}
\caption{Same as Figure~\ref{fig:wasp78_s4_transit}, but for the transit of HATS-24~b.}
\label{fig:hats24_s4_transit}
\end{figure}

\begin{figure}
\centering
\includegraphics[width=1.\hsize]{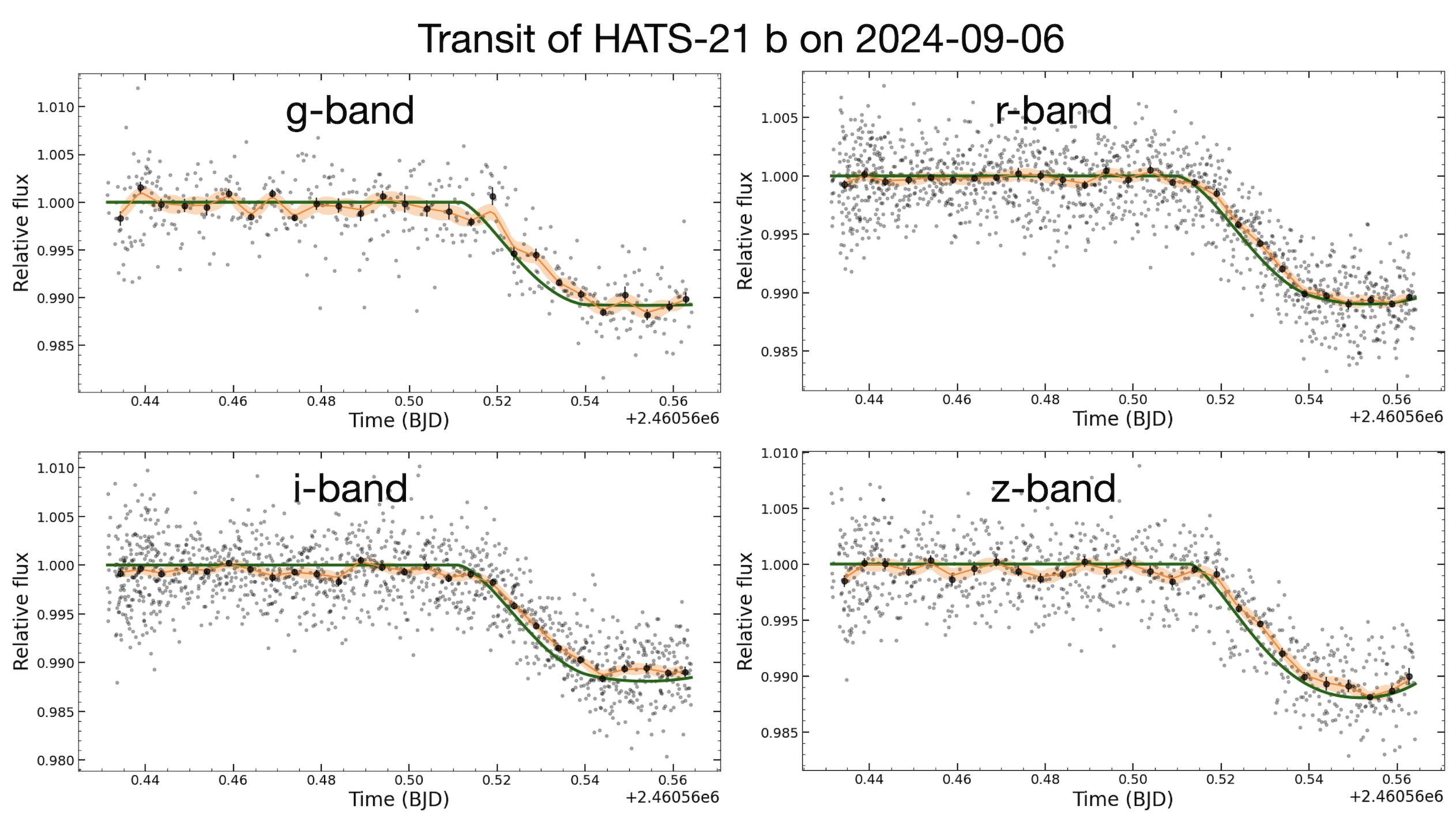}
\caption{Same as Figure~\ref{fig:wasp78_s4_transit}, but for the transit of HATS-21~b.}
\label{fig:hats21_s4_transit}
\end{figure}

\begin{figure}
\centering
\includegraphics[width=1.\hsize]{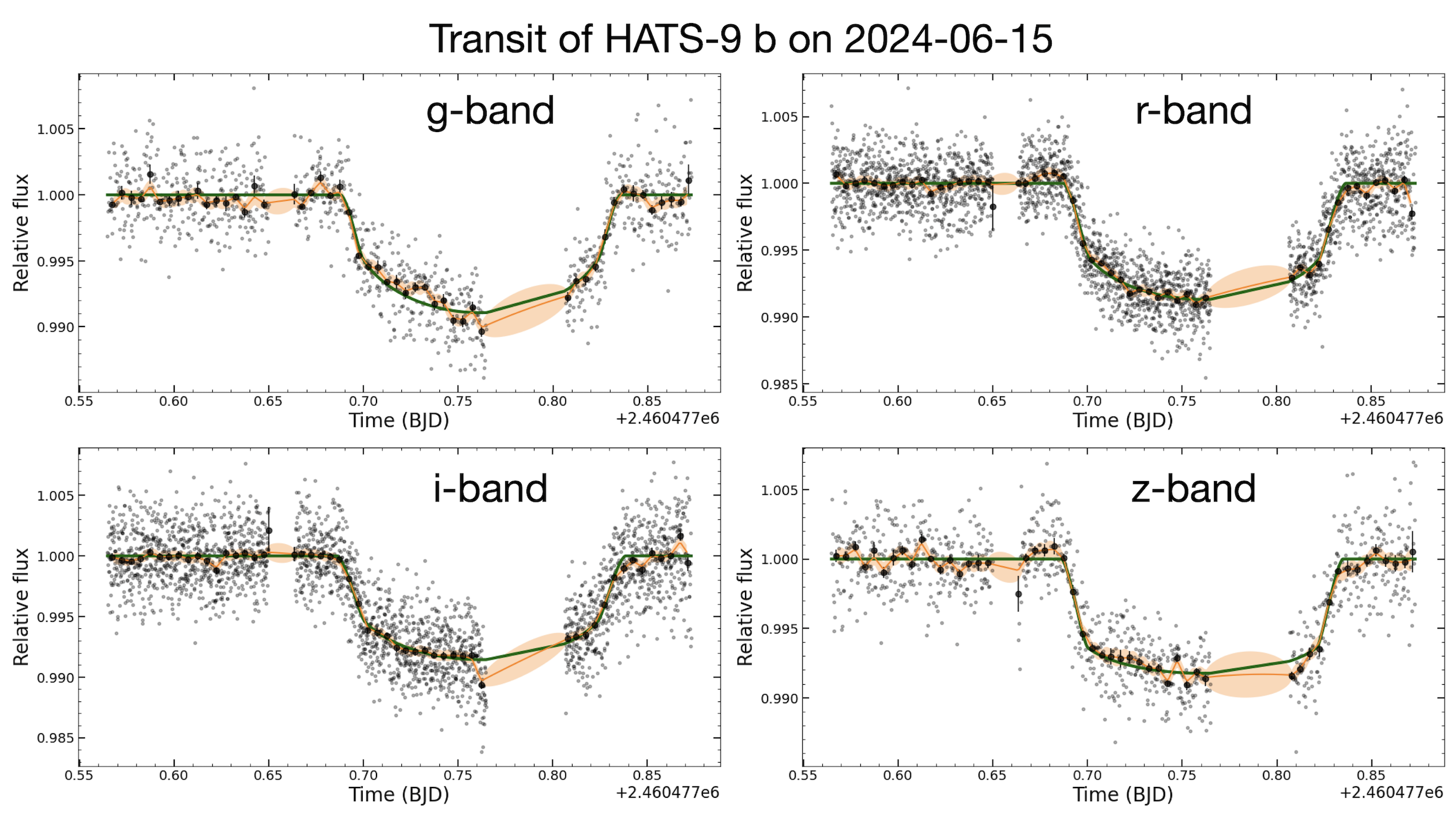}
\caption{Same as Figure~\ref{fig:wasp78_s4_transit}, but for the transit of HATS-9~b observed on 2024-06-15.}
\label{fig:hats9_s4_transit_2024-06-15}
\end{figure}

\begin{figure}
\centering
\includegraphics[width=1.\hsize]{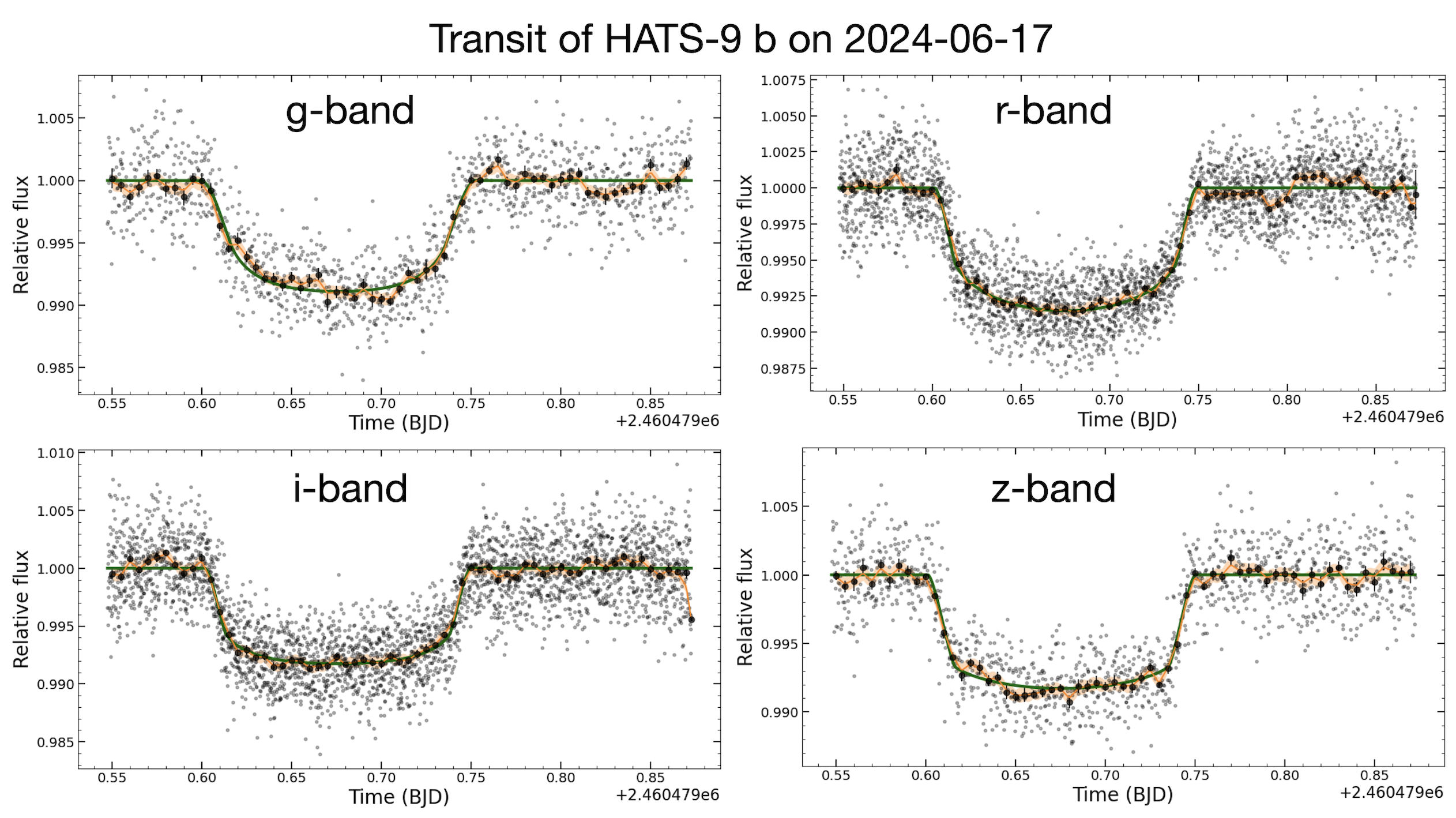}
\caption{Same as Figure~\ref{fig:wasp78_s4_transit}, but for the transit of HATS-9~b observed on 2024-06-17.}
\label{fig:hats9_s4_transit_2024-06-17}
\end{figure}

Figures~\ref{fig:wasp78_s4_transit}, ~\ref{fig:wasp123_s4_transit}, \ref{fig:wasp111_s4_transit}, \ref{fig:hats23_s4_transit}, \ref{fig:hats24_s4_transit}, \ref{fig:hats21_s4_transit}, \ref{fig:hats9_s4_transit_2024-06-15}, and \ref{fig:hats9_s4_transit_2024-06-17} show the final detrended four-band light-curve data for the eight transits of the seven exoplanets observed with SPARC4.

%% For this sample we use BibTeX plus aasjournalv7.bst to generate the

\bibliography{bibliography}{}
\bibliographystyle{aasjournal} %% nao mostra iniciais do primeiro autor



\end{document}